\documentclass[11pt]{article}
\newcommand{\insubfile}[1]{\ifx\@onlypreamble\@notprerr\else#1\fi}
\usepackage{xpatch}
\usepackage{hyperref}
\usepackage{geometry}
\usepackage[normalem]{ulem}
\usepackage{booktabs}
\usepackage{multirow, bigdelim}
\usepackage{graphicx}
\usepackage{subfiles}
\usepackage{pdfpages}
\graphicspath{ {figs/} {Extended/} }
\def\arcsec{{$^{\prime\prime}$}}

\newcommand{\aap}{    {\it Astron. Astrophys. }}
\newcommand{\aaps}{   {\it Astron. Astrophys. Suppl. }}

\newcommand{\apj}{    {\it Astrophys. J. }}
\newcommand{\apjl}{   {\it Astrophys. J. Lett. }}

\newcommand{\jfm}{    {\it J. Fluid Mech. }}

\newcommand{\nat}{    {\it Nature }}
\newcommand{\natcomm}{{\it Nat. Comm.}}

\newcommand{\pasj}{   {\it Pub. Astron. Soc. Japan }}
\newcommand{\pre}{    {\it Phys. Rev. E }}
\newcommand{\prl}{    {\it Phys. Rev. Lett. }}
\newcommand{\solphys}{{\it Solar Phys. }}
 
\newcommand{\ssr}{    {\it Space Sci. Rev. }}
\newcommand{\sci}{    {\it Science. }}

\newcommand{\apjs}    {{\it Astrophysical Journal Supplement Series}}

\newcommand{\pnas}{{\it PNAS}}

\title{Polymeric jets throw light on the origin and nature of the forest of solar spicules}
\date{}
\begin{document}
\maketitle
\author{{\centering{Sahel Dey$^{1,2}$, Piyali Chatterjee$^{1\ast}$, Murthy O. V. S. N.$^{3}$, Marianna B. Kors{\'o}s$^{4,7,8}$, Jiajia Liu,$^5$ \\ Christopher J. Nelson$^5$, Robertus Erd{\'e}lyi $^{6,7,8}$}\\
\vspace*{0.5cm}
\normalsize{$^1$Indian Institute of Astrophysics, Koramangala, Bangalore-560034, India}\\
\normalsize{$^2$Joint Astronomy Programme and Department of Physics, Indian Institute of Science, Bangalore-560012, India}\\
\normalsize{$^3$School of Arts and Sciences, Azim Premji University, Hosahalli Road, Bangalore-562125, India}\\
\normalsize{$^4$Department of Physics, Aberystwyth University\\
Ceredigion, Cymru SY23 3BZ, UK}\\
\normalsize{$^5$Astrophysics Research Centre, School of Mathematics and Physics, Queen's University Belfast, Belfast BT7 1NN, UK}\\
\normalsize{$^6$Solar Physics and Space Plasma Research Centre (SP2RC),  School of Mathematics and Statistics, University of Sheffield, Hicks Building, Hounsfield Road, Sheffield, S3 7RH, UK}\\
\normalsize{$^7$Department of Astronomy, E\"otv\"os Lor\'and University, P\'azm\'any P. s\'et\'any 1/A, Budapest, H-1117, Hungary}\\
\normalsize{$^8$Gyula Bay Zolt\'an Solar Observatory (GSO), Hungarian Solar Physics Foundation (HSPF), Pet\H{o}fi t\'er 3., Gyula, H-5700, Hungary}\\
\normalsize{$^\ast$To whom correspondence should be addressed; E-mail:  piyali.chatterjee@iiap.res.in}
}}

\baselineskip24pt
\maketitle 

\begin{abstract}
Spicules are plasma jets, observed in the dynamic interface region between the visible solar surface and the hot corona. At any given time, it is estimated that about 3 million spicules are present on the Sun. We find an intriguing parallel between the simulated spicular forest in a solar-like atmosphere and the numerous jets of polymeric fluids when both are subjected to harmonic forcing. In a radiative magnetohydrodynamic numerical simulation with sub-surface convection, solar global surface oscillations are excited similarly to those harmonic vibrations. The jets thus produced match remarkably well with the forests of spicules detected in observations of the Sun. Taken together, the numerical simulations of the Sun and the laboratory fluid dynamics experiments provide insights into the mechanism underlying the ubiquity of jets: the nonlinear focusing of quasi-periodic waves in anisotropic media of magnetized plasma as well as polymeric fluids under gravity is sufficient to generate a forest of spicules on the Sun.
\end{abstract}

Spicular jets are highly elongated [$4000-12000$ km in length, $300-1100$ km in width] features which are believed to transport momentum to the solar wind and non-thermal energy to heat the atmosphere\cite{Beckers72,Sterling00,Tsiropoula12}. 
Several mechanisms have been proposed to account for the formation of solar spicules, including granular squeezing\cite{Roberts79}, shocks and pulses\cite{Hollweg82,Suematsu82,Iijima15}, solar global acoustic waves\cite{DePontieu04,Heggland07,Heggland11}. So far, models based on these drivers have not been able to quantitatively match both the heights and abundance of the observed solar spicules. This is one of the reasons why the focus in the community has shifted to other mechanisms e.g., Alfv\'en waves\cite{Haerendel92, Liu19}, magnetic reconnection\cite{Sterling91,Shibata07,Samanta19}, magnetic tension release aided by ion-neutral coupling\cite{Martinez-Sykora17}, and the Lorentz force\cite{Iijima17} to produce a few jets whose heights agree better with observations.

A visually similar jetting behaviour is also seen at the free surface of a fluid layer when vertically oscillated in what is known as Faraday excitation\cite{Faraday1831}. Resonant mode coupling\cite{Rajchenbach13} in such systems leads to horizontal pattern formation. Beyond a threshold excitation amplitude, droplets or jet ejections occur by subsequent focusing of surface gravity-capillary waves 
in Newtonian fluids\cite{Goodridge96} as well as polymeric solutions\cite{Wagner99, Cabeza07}. These table-top experiments of jets in a buffeted polymeric fluid motivates the question: would not such a mechanism also assemble a forest of spicules in a solar simulation, where the anisotropic magnetized solar plasma is analogously buffeted by convection and global modes. Physically, the interaction of magnetic fields with the plasma turbulence is analogous to that between long-chain polymers and thermal fluctuations in a viscoelastic fluid\cite{Longcope03}. This parallel has been shown to exist in a shearing geometry (i.e., Couette flow) as the viscoelastic stresses in dilute polymeric fluids which follow the Oldroyd-B model and the Maxwell's stresses in magnetohydrodynamics (MHD) have mathematically similar off-diagonal terms\cite{Ogilvie03} that physically manifest as anisotropy in  the respective mediums. We extend the analogy between Oldroyd-B model and MHD to elongational stresses, $\sigma_{ij}$, where the magnetic field in the plasma plays a role analogous to polymeric fluid properties dependent on concentration . In a nutshell, we can compare the diagonal term or $\sigma_{zz}$ in the stress tensor due to uniaxial elongational stresses, in both systems, as $$\frac{B_x^2+B_z^2}{2\mu_0} \leftrightarrow \frac{2\mathrm{Tr}(c) \eta_p(c)\dot{\epsilon}}{3}.$$
Here, $\mu_0$ is the permeability,  $B_x$, $B_z$ are magnetic field components along the $x$ and $z$-directions, respectively. The Trouton ratio which is a function of polymer concentration is denoted $\mathrm{Tr}(c)$, the viscosity of the polymeric solution at no applied shear is $\eta_p(c)$, and $\dot{\epsilon}$ is the strain rate [see \S 1 of Supplementary Information for details]. In this work, we provide evidence of how this similarity can be utilized to arrive at the common conditions for the formation of a forest of jets in both the systems by subjecting the solar-like atmosphere to harmonic oscillations similar to Faraday excitation in fluids. The term "forest" is used here for several jets taller than a threshold height seen in both scenarios. The ejection locations do not have a strong relation to the shape of the container or the spatial pattern of the driver. It also refers to spicules obtained in a range of heights commensurate with solar observations. By implementing harmonic vibrations self-consistently generated by subsurface solar-like convection, we identify a model that is able to capture the features of the forest of solar spicules.

\section*{Phenomenological parallels}
To begin with, we perform an ensemble of 2-dimensional numerical experiments on the solar atmosphere using the Pencil code (see Methods), wherein a harmonic forcing is applied just below the photosphere. Curiously, this mimicking of the naturally occurring solar global oscillations generates a forest of spicules, in the presence of a vertically imposed magnetic field. The threshold photospheric acceleration, $A_\mathrm{min}$, above which jets satisfying a set of criteria are formed, is shown in Fig.~\ref{fig:phase_plot}a, varying with the driving frequency, $f_0$ (scaled to approximate minimum $f_s=4$\,milli Hertz, mHz). Similar protocols have been applied to arrive at the threshold value of acceleration for the harmonic driving experiments with plasma as well as the polymeric fluids in the relevant domain sizes, namely: i) at least 4 jets in ten cycles reaching a height $\ge 7$\,Mm from photosphere (plasma) and $\ge 0.5$\,cm above the static level (fluid) should be seen after the nonlinear development phase described later, and ii) the jets must have an aspect ratio (width:height) lower than 0.5. The chromospheric plasma experiences a much larger acceleration due to lower density. It may be noted that the region R2, around $f_s$, is also where the solar global oscillation power spectrum exhibits a peak\cite{Garcia05}. We also demonstrate spicule formation for a case with a quasi-periodic forcing instead of a perfectly harmonic driver in Supplementary Video~\ref{mov:mixed}. These findings suggest that the vertical bobbing of the photosphere is the significant driver of spicule formation\cite{Uchida61,Suematsu82,DePontieu04}. For comparison, we conduct a series of table-top experiments on a dilute polymeric solution of polyethylene oxide (PEO) excited using a speaker at low audio frequencies.
For forced vertical acceleration with a time dependence, $A_0 \sin(2\pi f_0 t)$, the $A_\mathrm{min}$ at each $f_0$, where the above forest of jets criteria is satisfied, is arrived at by carrying out a series of individual experiments by sampling the amplitude, $A_0$, corresponding to each $f_0$. This process, as an example, is visualized by indicative data points represented by filled and open symbols about the threshold in both the Figs.~\ref{fig:phase_plot}a-b (also see Methods for the protocol and \S4 of Supplementary Information for an example). At other frequencies, we represent the lowest acceleration amongst those runs where the forest condition is satisfied. The behaviour of the $A_\mathrm{min}-f_0$ curve in region R2 (see Fig.~\ref{fig:phase_plot}b) is similar in both cases. In the fluid, for low driving frequencies (i.e. $< f_s=25$\,Hz or region R1), the acceleration threshold for jets shows a flattening {\em w.r.t.} $f_0$.
This behavior agrees well with the existence of a cut-off forcing amplitude, obtained by solving the Mathieu equation, only above which nonlinear mode coupling is permissible\cite{Rajchenbach13}. Likewise, an increase (shown in region R1 of Fig.~\ref{fig:phase_plot} a) in threshold acceleration with decreasing frequency is also seen upon reducing $f_{0} < 4$ mHz likely due to the existence of the solar acoustic cut-off frequency in the lower solar atmosphere ($\sim 3.3$ mHz).
In the driving frequency range of 30--90\, Hz (region, R2), the $f_0^{4/3}$ dependence obtained for jets by dimensional arguments for a fluid under Faraday excitation\cite{Goodridge97} is also shown for comparison. 

The importance of long polymeric chains becomes evident at higher frequencies, where we note a change in slope at larger driving frequencies ($f_0>90$\,Hz or region R3). This indicates the region in $A_0-f_0$ phase space where the jet, as a result of dispersion relation of the surface gravity-capillary waves (see Methods), becomes thin enough for anisotropic elastic effects to dominate the dynamics also referred to as elasto-capillary regime\cite{Dinic19}. At higher frequencies (R3) the behaviour of the polymeric solution is different from that of the solar plasma (Fig.~\ref{fig:phase_plot}a). In contrast, a non-polymeric fluid (55\% glycerine solution with comparable viscosity as 1000\,ppm PEO solution) presents only droplet ejections and no coherent jets at these high frequencies (also see Supplementary Video~\ref{mov:jetbreaking}a). The role of anisotropy in the jet formation process is further illustrated by comparing the effect of the same excitation on water and a dilute PEO solution in Fig.~\ref{fig:phase_plot}d. 
The consistent and sharp decrease in the number of droplets observed, even at low polymer concentrations is shown in Fig~\ref{fig:phase_plot}f. The droplets are formed by pinching of jets by parasitic capillary waves via the Plateau-Rayleigh Instability\cite{Plateau1873, Rayleigh1892}. The measured fluid turbulence (see Methods) points to increasing energy absorption by the uncoiling of the polymer chains and also correlates quite well with the droplet count. This stretching of the polymer chains along the jets may be imaged using a crossed polarizer setup (Extended Data Fig.~\ref{fig:pol_setup} with iodinated Poly vinyl alcohol (PVA) as the polymeric solution instead of PEO). The list of the solar plasma runs and polymeric fluid experiments is provided in Extended Data Table~\ref{tab:tab2}.

To investigate the role of magnetic field in the vertically oscillated solar plasma, we set the magnetic field to zero in the simulations. We then observe only small-scale Kelvin-Helmholtz (KH) vortex plumes instead of jets, as shown in Fig.~\ref{fig:phase_plot}c. Interestingly, anisotropy created in an artificial manner by introducing a horizontal velocity damping at zero field also results in the reappearance of jets in the undamped vertical direction (see also Extended Data Fig.~\ref{fig:zero_mag}). It may therefore be concluded that the magnetic field provides this anisotropy by collimating the rising plasma to form jet-like structures via the Maxwell's stress tensor, thereby suppressing the KH instability of the rising plumes. The effect of imposing even a weak magnetic field, say $B_\mathrm{imp}\sim 1$\,G, in the plasma is analogous to addition of a low concentration (50\,ppm) of polymer in a low viscosity solvent (Fig.~\ref{fig:phase_plot}e, f); both provide anisotropy, key to vertical jetting behaviour in the presence of large quasi-periodic forcing (also see \S2 \& 3 of Supplementary Information for an explanation of the existence of critical anisotropy measures beyond which KH and Plateau-Rayleigh instabilities are inhibited, respectively). 

The phenomenological parallels also extend to a few non-dimensional quantities (e.g. aspect ratio, Froude number, etc.; see Extended Data Table~\ref{tab:tab1}) as well as the nonlinear dynamic evolution from the start of the vertical excitation of the solar plasma in a 3-dimensional box unlike a 2-dimensional domain used previously (see Methods for details of harmonic forcing) and in the polymeric 2\% PVA solution (see Supplementary Video~\ref{mov:plasma_fluid}). The solar plasma in the 3D numerical experiment follows the spatial dependence of the forcing for a few periods before breaking up into a crispate (or corrugated) structure. Whilst in the case of the Faraday-excited fluid, initially contra-propagating axisymmetric wave crests travel to-and-fro forming an inertial-focusing driven central jet\cite{Zeff00}. Several jets are then formed from the circular wave crests which now resemble a crispate structure. Afterwards, we see the appearance of sub-harmonic surface distortions resembling polygonal cells. Jets are usually ejected from regions where the surface convolutions induced by Faraday excitation become focused due to the collision of ridges of neighboring polygonal cells (see Supplementary Video~\ref{mov:mechanism}). A similar forest of PVA jets also forms when instead of a purely harmonic excitation, an accelerated quasi-periodic solar acoustic excitation is used (p-modes, also Supplementary Video~\ref{mov:pmode}). 

\section*{Solar spicule forest with convection}
In order to demonstrate the utility of this model for solar spicules,  
we introduce a solar-like convective layer below the photosphere. The computational results from one representative simulation run performed in a 2D box is visualized using synthetic emission from the plasma at a temperature of 80000 K (Fig.~\ref{fig:sun_lab}b, also see Supplementary Video~\ref{mov:SiIV}), which portrays the upper chromosphere, and using the temperature response function of the $17.1$ nm filter of the Solar Dynamics Observatory's Atmospheric Imaging Assembly\cite{Lemen12} (SDO/AIA), which maps the lower solar corona (Fig.~\ref{fig:sun_lab}c). A typical forest of observed jets is shown in Si IV $140$\,nm (plasma temperature $\sim 80000$\,K) images sampled at the solar limb by the Interface Region Imaging Spectrograph--IRIS\cite{DePontieu14} (see Fig.~\ref{fig:sun_lab}a where an example spicule is denoted by the arrow labelled `S1' and Extended Fig.~\ref{fig:MgII}a for a wider field of view). The simulated forest of spicules (panel b) qualitatively agrees well with the spicule forest from observations (panel a). The simulated spicule heights, as measured by synthetic emission peaked at a temperature of 15000\,K (see Extended Data Fig.~\ref{fig:MgII}d), vary between $6-24.6$ megameters (Mm) with an average of $12.6$\,Mm and maximum upward velocities between $30-85$ km s$^{-1}$, quantitatively agreeing with the statistics of observed spicules\cite{Tsiropoula12,Pereira14} (Extended Data Fig.~\ref{fig:MgII}b). The time-distance maps in the right panels highlight the parabolic nature of the observed\cite{Anan10} and simulated spicules, as well as fluid jets in acrylic paint (also, Supplementary Video~\ref{mov:paint}). The measured deceleration is consistently higher for polymeric jets as compared to non-polymeric ones. To illustrate this, we use five isolated and vertical representative jets in 1000\,ppm PEO solution ($29.5 \pm 4.4$\,m\,s$^{-2}$) as well as glycerine solutions ($15.8\pm 1.6$\,m\,s$^{-2}$) as compared to ejected droplets ($9.5\pm0.7$\,m\,s$^{-2}$) (see Supplementary Video~\ref{mov:jetbreaking}b at 30\,Hz for few exemplary cases). This may indicate the additional tension forces exerted by the uncoiled polymer chains in PEO for polymeric jets in contrast to glycerine solution of similar viscosity.  

We now show that this model of solar spicule generation brings out the following insights into the origin and the structure of observed spicules. Using the technique of Lagrangian tracking, it is clear that inside a spicular structure the plasma at the photospheric and low chromospheric heights does not rise all the way to corona (see Supplementary Video~\ref{mov:tracking}a). All the plasma that rises to coronal heights ($> 6$ Mm) comes from above $2$ Mm. The disparity in the heights reached by plasma originating from separate layers potentially explains the varying lengths of spicules in different observational channels\cite{Pereira14}. 
The numerical experiment finds that even if the upward acceleration measured in the chromosphere is only $\sim 4 g_\mathrm{S}$, in terms of the gravity at the solar surface but reaching  as large as $12 g_\mathrm{S}$ in some regions between 1--2\,Mm, the plasma is energized to speeds above 30\,km\,s$^{-1}$ (Fig.~\ref{fig:layer_track}d) by an acceleration front propagating ahead and upwards at the speed of sound (Extended Data Fig.~\ref{fig:Ma_shock}, and Fig.~\ref{fig:spicheat}a). Because of this front, the entire plasma inside the synthetic solar spicule does not rise and fall in unison. Usually, the plasma inside the front is still speeding upwards even while the plasma below the front is either slowing its rise or is falling down (Supplementary Video~\ref{mov:tracking}b--c). Obviously, the acceleration front communicates the dominant photospheric acoustic forcing to the rest of the atmosphere and is also the region of large compression or shock. This front exhibits steepening as it propagates into the region of decreasing density from the lower chromosphere to the corona. The heating inside this front is dominated by compression of the shocked plasma rather than dissipation due to viscosity or Ohmic currents as evident from the Extended Data Fig.~\ref{fig:spicheat}a--c. 
The technique of Lagrangian tracking brings to the fore the fine structure of the flows inside the spicules; the detection of such flows within spicules may be possible using the ultra-high resolution observations of the solar chromosphere that will be revealed by the next generation of solar telescopes.

We report a continuously decaying distribution of spicule heights in our simulation as well as fluid jets in the experiment with shorter being dominant in number than longer jets (see Extended Data Fig.~\ref{fig:hist}a, b). Note the slightly enhanced abundance at the tall end of the height distribution. The $p$-value from the Kolmogorov-Smirnov statistical test on the cumulative distributions is 0.41 (Extended Data Fig.~\ref{fig:hist}c), indicating that the null hypothesis of both distributions being similar cannot be rejected. At least two classes of spicules have also been reported in observations\cite{Beckers72,DePontieu07}. In this context, our findings are summarized below.

\emph{Short spicules}, in our simulations, form above the convective down-flow regions where the magnetic field opens up as a funnel. The formation of a convective plume squeezes the magnetic flux tube and, therefore, forces the plasma trapped inside to shoot upwards along the field lines into the upper atmosphere (Fig.~\ref{fig:layer_track}b). The convergence speed at the photosphere is around 1\,km\,s$^{-1}$ as estimated\cite{Roberts79}. 
Additionally, we also report an event where a collapsing granule between two convective plumes (P1 and P2) causes these plumes to collide and eject a spicule (see Supplementary Video~\ref{mov:tracking}d).
\emph{Long spicules} form in regions above the convective granules where the magnetic field is organized in the form of low-lying loops. The sub-surface convection excites motions whose power spectrum is peaked at about 3.3\,mHz (solar global oscillations) causing the solar surface, represented by the boundary of the convective granules, to rise and fall quasi-periodically. 
This leads to formation of a stronger acceleration front than those above the squeezing sites as shown in the phase plots of vertical components of acceleration and velocity, namely, $a_z$ vs $v_z$ in Extended Data Fig.~\ref{fig:azvz_track_Lorentz}b as compared to panel (d). It has been argued that the power of the solar global oscillations reaching the atmosphere is usually higher over a granule than above an inter-granular lane\cite{Hansteen06, DePontieu07}. In the chromosphere, the resultant plasma motions push oppositely directed magnetic field lines together causing magnetic reconnection between the open field lines and the low lying loops. This process aids the  plasma to escape along the newly opened field lines leading to the generation of faster, and longer spicules as shown in Figs.~\ref{fig:layer_track}c--d (also see Supplementary Video~\ref{mov:tracking}e).
Additionally, long spicules display bright tips when the AIA $17.1$ nm intensity is synthesised, (blue circles over-laid on Fig.~\ref{fig:sun_lab}c, also Supplementary Video~\ref{mov:SiIV}), consistent with observations\cite{Pereira14,Henriques16,Samanta19}. 

The enhanced emission can be further seen in the time-distance maps of both observed and simulated spicules (Fig.~\ref{fig:sun_lab}a--c) whereas the shorter spicules do not have discernible emission at the tip. In either case, the vertical acceleration due to the Lorentz force does not appear to play a major role in accelerating the plasma here (Extended Data Fig.~\ref{fig:azvz_track_Lorentz}c and f).

In terms of underlining physics of jetting phenomena in the two systems, the commonalities are: i) a large amplitude quasi-periodic forcing leads to nonlinear steepening of outward propagating acceleration fronts (in compressible and stratified plasma) or focusing of surface gravity-capillary wave crests (incompressible fluid); ii) anisotropic stresses in the solar atmosphere collimate the plasma and suppress the Kelvin-Helmholtz (KH) instability while the same in a buffeted polymeric fluid suppress the Plateau-Rayleigh instability;
iii) the apex height of the jets vary as a parabolic function of time; and, finally iv) a decaying distribution of heights exists for both the plasma and fluid jets. 
Despite the behavioural similarities significant differences between the two systems exist, compressibility and the relevant nature of tension being noteworthy. For the compressible solar plasma, the dominant power is in a narrow band around 5-min acoustic $p$-mode whereas the polymeric fluid does not have such modes at 30-120 Hz driving frequencies. However, the fluid system can still respond to an oscillatory driver by means of surface gravity-capillary waves.

\section*{Insights into the spicule forest}
In the context of solar spicules, we now report the following results, compared to previous works\cite{Iijima15,Martinez-Sykora17,Iijima17}. Namely, a forest of spicules are formed in our simulations with heights ranging between 6--25\, Mm, bearing significantly closer resemblance to clusters of jets observed in the solar atmosphere. The plasma from the chromosphere, in contrast to the heavier photospheric plasma, can be readily energised by acceleration fronts to form taller as well as shorter spicules. The acceleration fronts are themselves generated by several mechanisms including e.g. i) squeezing by granular buffeting\cite{Roberts79}, ii) collapse of granules\cite{Martinez-Sykora09} and, iii) solar global modes\cite{DePontieu04}, aided by magnetic reconnection, each ultimately induced by the same agent: convection. Are other mechanisms important? Specific mechanisms may be relevant for some kinds of spicules though (ambipolar diffusion\cite{Martinez-Sykora17}, Lorentz force\cite{Iijima17}, Alfven pulses\cite{Haerendel92,Liu19,Sakaue20}, magnetic reconnection\cite{Sterling91,Shibata07} etc). A question to explore would be whether these other mechanisms on their own are able to account for the abundance of the rapid and tall spicules as seen in this study, as well as in the Hinode\cite{DePontieu07} and BBSO data\cite{Samanta19}. 
In addition to the forest feature of spicules, our model also captures their main characteristics in spite of not including chromospheric microphysics of ambipolar diffusion and non-local thermodynamic equilibrium of the partially ionized  plasma\cite{Martinez-Sykora17}. 
The phenomenological and mathematical similarities between the fluid and plasma despite the lack of charge-related effects (e.g., Hall term, ion-neutral collision) as well as specific thermodynamic considerations in the fluid point to an even more fundamental cause behind the generation of a forest of spicules. 
The insight provided by the polymeric fluid experiments when combined with the commonalities with the numerical solar MHD simulations is that four basic ingredients - a fluid medium, gravity, large amplitude quasi-periodic driving, and anisotropy of the medium - are sufficient to assemble a forest of spicules on the Sun by nonlinear development.

 \subsection*{Acknowledgement} 
 Computing time provided by Nova HPC at IIA as well as Param Yukti facilty at JNCASR under National Supercomputing Mission, Govt. of India is gratefully acknowledged. This work has also utilized SahasraT facility at SERC, IISc for computing and CENSe, IISc for rheometry. 
 S.D. and P.C. thank IIA for financial support towards the usage of SahasraT. S.D. also acknowledges the SOLARNET project that has received funding from the European Union's Horizon 2020 research and innovation programme under grant agreement no 824135. M.O.V.S.N acknowledges a grant RC00226 to study Faraday oscillations in fluids from the Azim Premji University. M.B.K. is grateful to the Science and Technology Facilities Council (STFC) for the ST/S000518/1 grant. J.L. acknowledges the support from the Leverhulme Trust via grant RPG-2019-371. C.J.N. is thankful to STFC for the support received to conduct this research through grant numbers ST/P000304/1 \&  ST/T00021X/1. R.E. is grateful to STFC (grant No. ST/M000826/1), the Royal Society, and the President's International Fellowship Initiative of the Chinese Academy of Sciences (grant No. 2019VMA052) for enabling this research. R.E. also acknowledges the support received from the Higher Education Programme of Excellence for Particle and Astrophysics, E\"otv\"os L. University (Budapest, Hungary).  We are grateful to Chirag Kalelkar for several discussions on polymeric fluids. IRIS is a NASA small explorer mission developed and operated by LMSAL with major contributions to downlink communications funded by ESA and the Norwegian Space Centre. 
 The accelerated solar global $p$-mode observations used to construct Supplementary Video~\ref{mov:pmode} was obtained by the SoHO/MDI instrument. We have used the visualisation software Paraview for volume rendering for Supplementary Video~3.

\subsection*{Author Contribution} 
P.C. and M. O. V. S. N conceptualized the study. P.C. prepared the numerical simulation set-up of the solar atmosphere. P.C and S.D performed the simulations, analysed the results and prepared the figures and animations with contributions from J.L on ASDA and spicule statistics and suggestions from R.E. The fluid experiments were designed and conducted by M.O.V.S.N.  The experimental data was analysed by M.O.V.S.N with contributions from P.C., S.D and M.B.K. The IRIS data analysis was performed by M.B.K and R.E who also prepared the corresponding figures with contributions from C.J.N. and J.L. The $p$-mode time series was provided by R.E. The manuscript was drafted by P.C. and M.O.V.S.N with inputs from R.E. All authors contributed to the interpretation of the results and worked on the subsequent versions of the manuscript together.
 
 \subsection*{Competing Interests} 
 The authors declare that they have no competing financial or non-financial interests.

\begin{figure}
\includegraphics[width=0.95\textwidth]{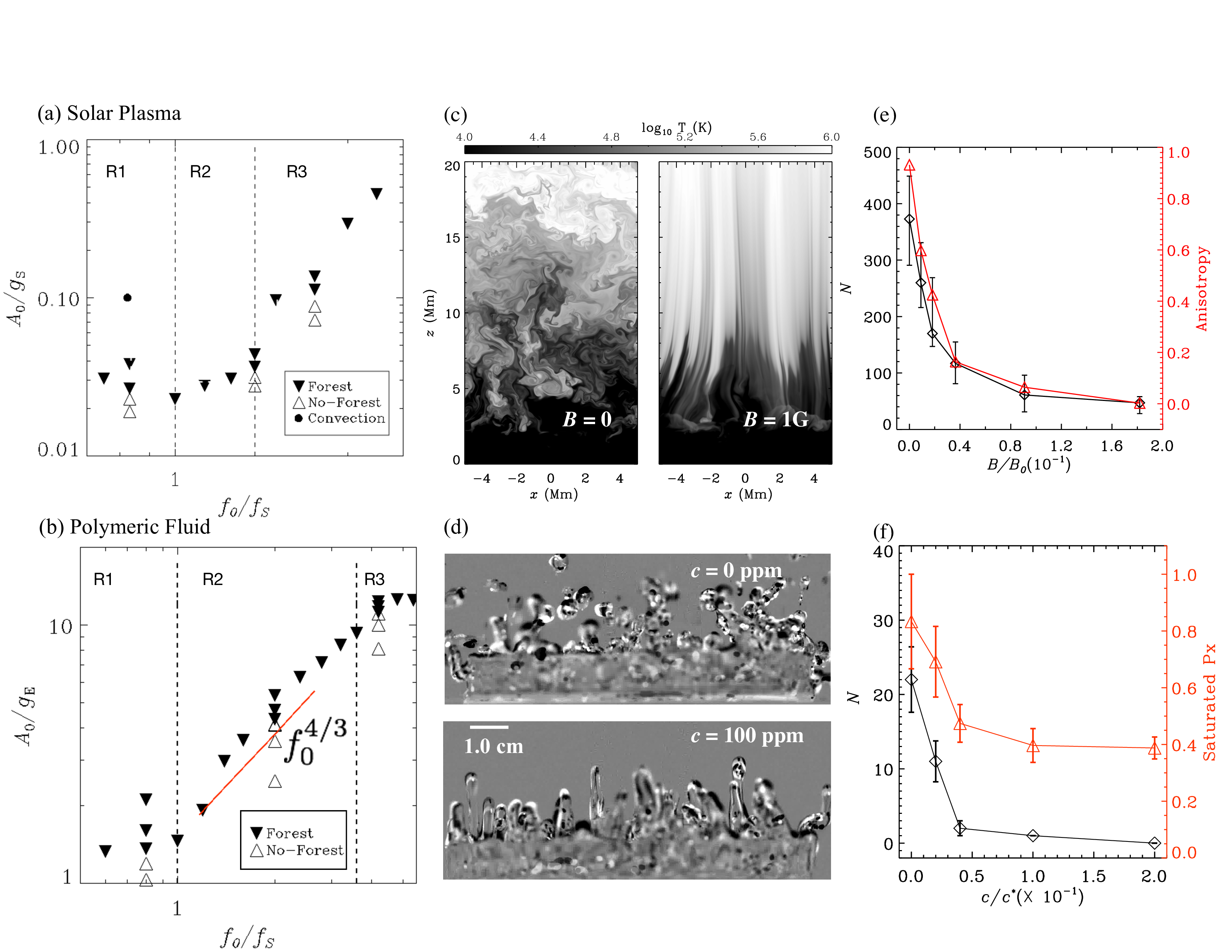}
\caption{\label{fig:phase_plot} {\bf Large acceleration and anisotropy}: (a) An $A_0$ versus $f_0$ phase plot to determine the threshold acceleration, $A_\mathrm{min}$, at which the forest of jets criteria is satisfied. Each data point represents an independent simulation run, in the $A_0/g_\mathrm{S}$ versus $f_0/f_s$ space with $f_s=4$\,mHz, for the solar atmosphere subjected to a harmonic driving at frequency $f_0$. Any simulation with given $A_0$ and $f_0$ either produces a forest of jets (filled triangle) when above a threshold or it does not (open triangle) when below. The dashed vertical lines are used to mark the regions R1, R2, and R3. The filled circle denotes the position of the 2D run with solar convection in this phase space. (b) Same as (a) but for polymeric fluid (500 ppm PEO) with $f_s=25$\,Hz and acceleration scaled by gravity, $g_\mathrm{E}$. Here, also, each triangle denotes an independent experimental run. The red guiding line denotes the expected dimensional dependence $f_0^{4/3}$ for fluids. (c) Snapshots of the simulation with a harmonic forcing of amplitude 3.7 times larger than the threshold at 3.3\,mHz without and with an imposed magnetic field $B_\mathrm{imp}=1$\,G, respectively, as indicated. (d) Difference images for water and PEO (100\,ppm), respectively at a driving frequency $f_0=30$\,Hz and peak acceleration of 10 $g_\mathrm{E}$. (e) Effect of $B_\mathrm{imp}$ on the average number of vortices detected over five snapshots equally sampled within a 5\,min oscillation time period by Automated Swirl Detection Algorithm (see Methods), where the equipartition field at $z=1.05$\,Mm is $B_0=5.5$\,G, for the same forcing as used in (c) (open diamond). Anisotropy in the images (open triangle) derived from distribution of the eigenvalues of the anisotropy gradient matrix. (f) Effect of polymer concentration, scaled by $c^*=500$\,ppm on number of droplet ejections (open diamond) and a measure of turbulence by counting saturated pixels or reflections (open triangle) obtained by averaging the maximum number of droplets/saturated pixels per cycle (18 frames) over five vibration time periods. The error bars denote the standard deviation within the same image frames mentioned in (e) and (f), respectively.}
\end{figure}

\begin{figure}
\includegraphics[width=0.43\textwidth]{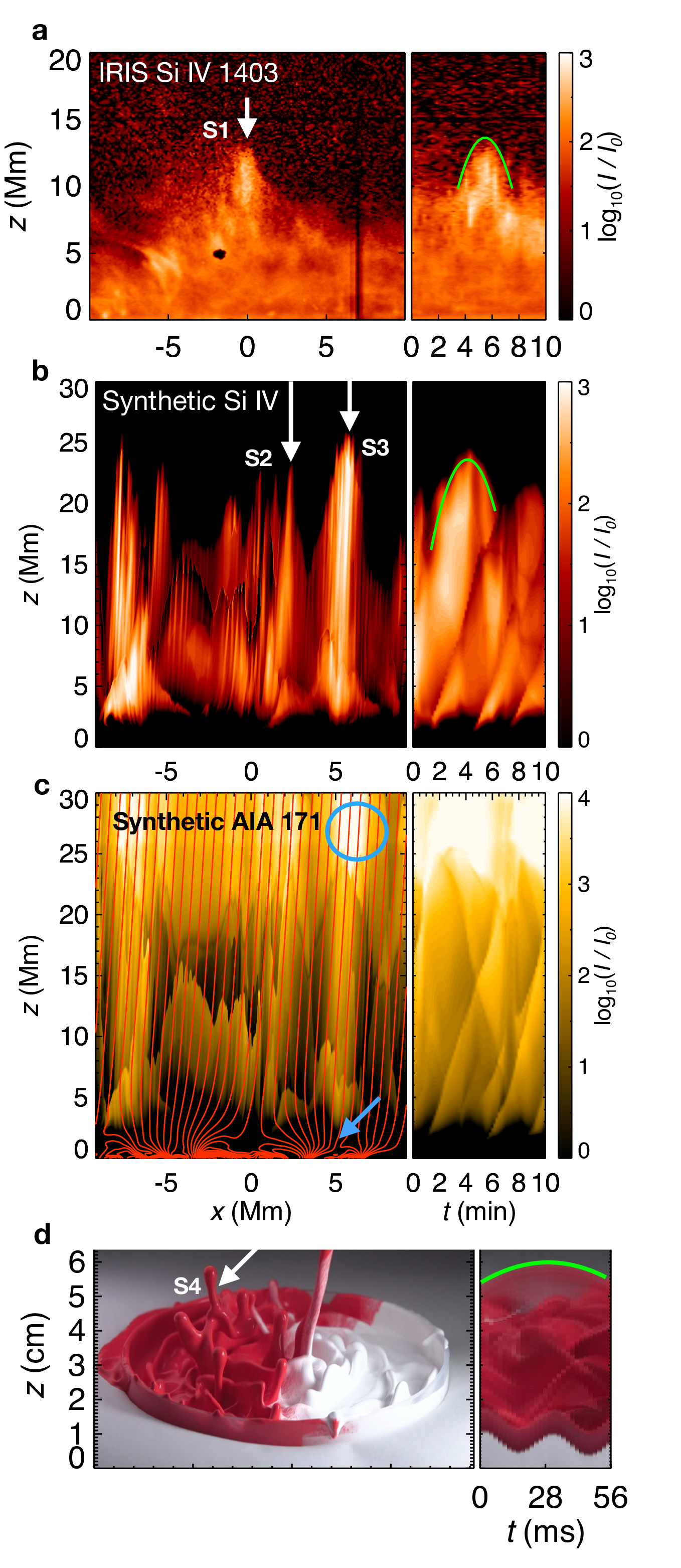}
\caption{\label{fig:sun_lab}  {\bf Comparison of 2D simulation with observation and laboratory experiment.} Left panels: (a) Spicules seen at the solar limb with the Si IV filter of IRIS. (b) Shaded contours of synthetic intensity for plasma at $T = 8\times 10^4$\,K (using Eq.~\ref{eq:syn}), for the 2D simulation run with $B_\mathrm{imp}=74$ G and solar-like convection at time, $t=152.50$ min from the start are shown in the $[x,z]$--plane. (c) Same as (b) but for the synthetic intensity of the AIA 17.1\,nm spectral line. The blue circle denote the region of increased emission intensity in extreme ultraviolet 17.1 nm line above the spicule captured by the slit, S3. The blue arrow points to the magnetic reconnection occurring below and the solid red lines represent the magnetic field.  The intensities are scaled by their horizontal average denoted $I_0 (z)$ at every height.} (d) Jets of acrylic paint for Faraday excitation at 30\,Hz with an amplitude of $7 g_\mathrm{E}$. Right panels: (a)--(d) corresponding time-distance plots for the slits S1--S4 indicated. The green curves in (a), (b) and (d) are the parabolic fit to the time-distance plots for isolated jets S1, S2 and S4. 
\end{figure}

\begin{figure}
\includegraphics[width=0.40\textwidth]{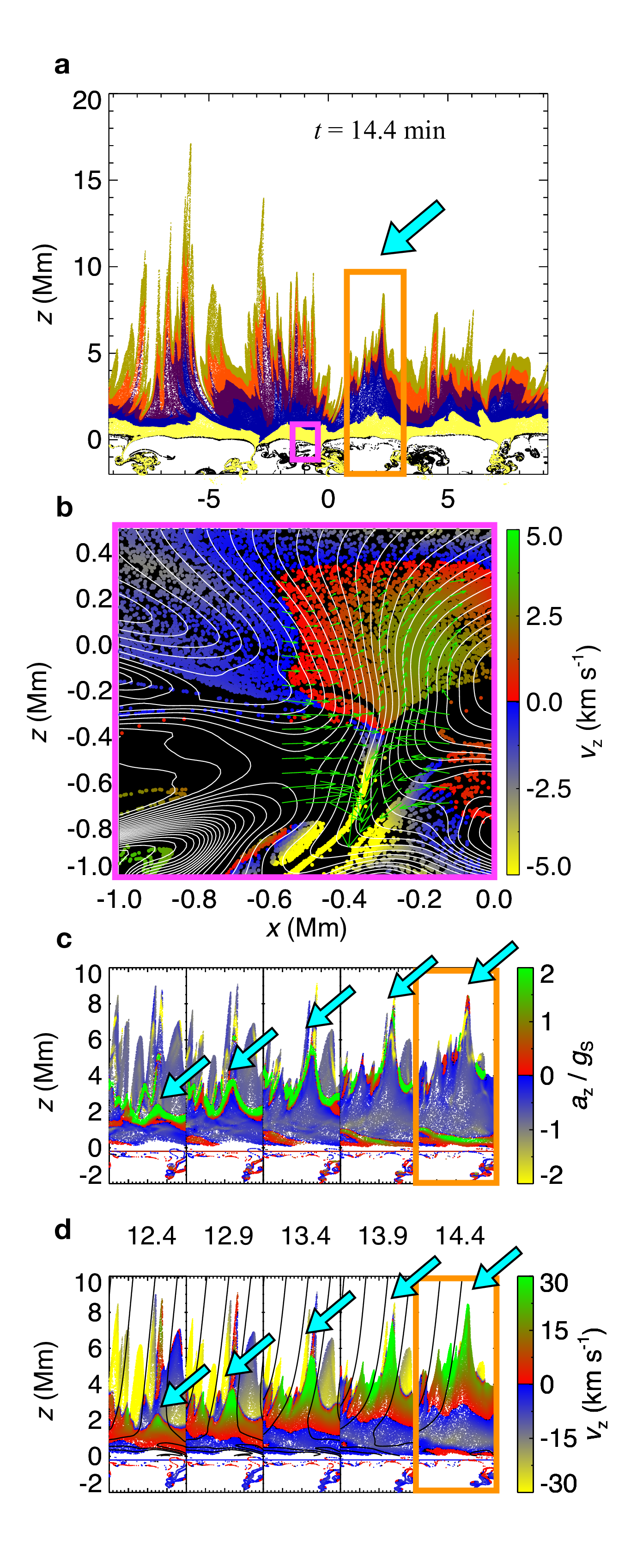}
\caption{\label{fig:layer_track}  {\bf Origin of forest of spicular plasma in the 2D simulation with convection}: (a) Position of 540000 Lagrangian tracking particles initially arranged in six different vertical layers (indicated by different colors) at 14.4 min from the time the particles were first introduced. (b) Magnetic field lines (white) and tracking particles (colored by vertical velocity) in the region of convective downdraft shown by the magenta box in panel (a). The green arrows indicate the velocity field in the region where formation of the plume squeezes the magnetic flux tube. (c) A temporal sequence of the tracking particles along a rising spicule indicated by the turquoise arrow and bounded by the orange box in (a) (in a sub domain with $1<x<3$ Mm) are shown with the
particles shaded by the acceleration experienced relative to the solar surface gravity, 274 m s$^{-2}$. (d) Same as (c) except now the shading is by vertical velocity (in km s$^{-1}$). The time stamp of each plot between 12.4--14.4 min since the start of tracking is indicated.  The solid (black) lines correspond to the magnetic field lines in the domain. The turquoise arrows in (b) and (c) follow evolution of a particular spicule.} 
\end{figure}

\newcounter{mybibstartvalue}
\setcounter{mybibstartvalue}{40}
\xpatchcmd{\thebibliography}{%
  \usecounter{enumiv}%
}{%
  \usecounter{enumiv}%
  \setcounter{enumiv}{\value{mybibstartvalue}}%
}{}{}

\section*{Methods}
\subsection*{The radiative MHD set-up:}

In this work, we model the solar plasma using the single fluid MHD approximation, where the magnetic field, $\mathbf{B}$ in the lab frame is generated due to the plasma moving at non-relativistic velocity, $\mathbf{v}$. Following the Ohm's law the electric field, $\mathbf{E}$ can be related to the current density, $\mathbf{J}$ as: $\mu_0\eta \mathbf{J} = \mathbf{E}+\mathbf{v}\times \mathbf{B}$, where, $\eta$ is the resistivity of the plasma and $\mu_0$ is the magnetic permeability in vacuum. Our simulation set-up couples the sub-surface convection to the solar corona using (a) an equation of state with ionization fraction calculated from the Saha ionization formula assuming local thermodynamic equilibrium 
(LTE), (b) detailed radiative transfer 
equation which is solved by the method of long characteristics\cite{Heinemann06}. We assume the grey approximation with the source 
function given by the Planck's black body function, integrated over all frequencies.  Further, we use a Rosseland mean of bound-free, free-free opacities (in Kramer's law form) 
and H$^{-}$ opacity in the 
chromosphere\cite{Chatterjee20}. An optically thin radiative cooling\cite{Cook89} obtained by using a linear piecewise interpolation in $\log T$ is also used in the corona, (c) large anisotropic thermal conduction along magnetic field lines, and (d) semi-relativistic Boris correction 
to the Lorentz force. The use of Boris 
correction to the Lorentz force allows us 
to work with a time-step, $\delta t > dx/v_A$, where, $dx$ denotes the grid size and, 
$v_A$ is the local Alfv\'en speed\cite{Chatterjee20}. 
In the domain, we impose a constant vertical magnetic field, $B_\mathrm{imp}$ at all times so that the net magnetic flux out of the photosphere is positive. The role of $B_\mathrm{imp}$ is to emulate the large-scale dynamo generated poloidal magnetic field of the Sun which cannot be produced self-consistently in a box-shaped model with a shallow convection region and without any shear. The convection acts on this imposed magnetic field to generate the small-scale magnetic flux at the photosphere. The grid size is 16 km which is sufficient to resolve the thin transition region of the solar atmosphere. The top 12\,Mm of the vertical extent consists of a ``sponge" layer, the purpose of which is to absorb the outgoing waves without letting them reflect back into the domain. Right at the base of this sponge layer lies a "hot plate" which maintains the temperature at $T_\mathrm{SL}$. The base of the sponge layer is located at $z=32$ Mm where the temperature is maintained at $T_\mathrm{SL}=1$ million K. Since we do not have any explicit heat source in the corona, this method also prevents the corona from cooling during the time scales required for the convection to reach steady-state at $t\approx 60$ min.  

For quantitative comparisons with the fluid experiments, we use a 2-dimensional domain of width, $L=9$,\,Mm and extending vertically from $-0.4$\, Mm$< z < 44$\,Mm, with $z=0$ representing the photosphere. The initial stratification of temperature is obtained by collating the Model S\cite{Christensen96} for the interior and an atmospheric model\cite{Vernazza81}. The initial density corresponding to this temperature is obtained by solving the hydrostatic balance subjected to the ionized ideal gas equation of state with ionization fraction given by the Saha-ionization formula\cite{Chatterjee20}. We fix the temperature and density at the bottom boundary ($z=-0.4$\,Mm) to values taken from the Model S. The flow at the bottom boundary is stress free and the top boundary is open to flows. At the start of the simulation, $B_\mathrm{imp}=0$ and is increased to $12$\,G over a duration of $t_\mathrm{fin}=10$\,min solar time in the manner $B_\mathrm{imp}=B^0_\mathrm{imp}\sin^2(\pi t/2t_\mathrm{fin})$. This is so that the atmosphere remains in quasi magneto-hydrostatic equilibrium throughout. Thereafter, we drive this set-up with a periodic membrane forcing velocity of the form $V_0 \sin(2\pi f_0 t)\cos(n\pi x/L)$ within a subsurface layer, $-100<z<0$\,km, to investigate the role of artificial harmonic forcing. Here, $n$ is the number of crests of the membrane forcing that fits into length $L$ of the box (see Extended Data Table~\ref{tab:tab2} for values used for different runs). Below this layer, the velocities are set to zero. The amplitude of the corresponding acceleration is $A_0=2\pi f_0 V_0$. We have followed the above protocol to make sure that the excitation amplitude, uncontaminated by convective motions, can be clearly determined for the $A_0$ vs $f_0$ phase space. The horizontal wave number of five is chosen to agree with a typical solar granulation scale of 2\,Mm. To generate Fig.~\ref{fig:phase_plot}a,  the code is first run for 25 cycle periods ($1/f_0$), after the start of the wave driving, to check for nonlinear development. In the case when we observe several spicules close together formed by crispation ({\em like of a compound leaf}) of the wave front, we continue to run it for another 10 cycle periods to further check for the height and aspect ratio criteria. But, if we do not observe any such nonlinear development during the first 25 cycles, we discard the run and change the amplitude of driving suitably before conducting a new run.
For the left panel of the Supplementary Video~\ref{mov:plasma_fluid}, a three-dimensional box is used for comparison with the fluid set-up to show the parallels in nonlinear development. A very shallow sub-surface layer is used here as well (300\,km) but, with a grid size of 24\,km. A slightly different method compared to the 2D runs of Fig.~\ref{fig:phase_plot} has been followed in order to match the time to formation of forest of jets with that of the polymeric solution (in terms of $1/f_0$). Here, the membrane forcing is of the form $V_0 \sin(2\pi f_0 t)\cos(\pi x/L) \cos(\pi y/L)$, with $f_0=3.3$\,mHz and $V_0=1.32$\,km\,s$^{-1}$, and only applied within a layer, $-100$\,km $<z<0$\,km. In the 3D box, we do not set the velocities to zero below that layer (unlike in Fig.~\ref{fig:phase_plot}), forcefully. Therefore, we had to verify that if the membrane forcing is not applied, jets do not form within the duration of the animation (25\,min solar time). The root mean square (rms) velocity reached in this shallow layer at the end of 25\,min, in absence of additional harmonic forcing, is $\sim 0.4$\,km\,s$^{-1}$ which does not drive jets in this set-up. 

For comparison with solar observations, we use a two-dimensional model with a deeper (5\,Mm) statistically relaxed convective layer, so that the vertical extent is now $-5$\,Mm$< z < 44$\,Mm. We follow the same steps as above to arrive at the initial stratification of temperature and density. However, now the slopes, 
$d \ln \rho/dz$ and $d \ln T /dz$, in the ghost zones are set to values expected from hydrostatic balance at the vertical boundaries. The $B_\mathrm{imp}$ is increased as described previously from zero to $74$\,G over a duration of 60\,min solar time, after which the convection is allowed to relax for another 60\,min. The final magnetic field value of 74\,G has been chosen to agree with recent observations\cite{Kriginsky20}. The temperature, density and vertical velocity structure in the $[x,z]$-plane of the 2D run shown in Fig.~\ref{fig:layer_track} at $t=85$\,min from the start of the simulation, and also at the moment when Lagrangian particles are introduced is shown in the Extended Data Fig.~\ref{fig:convection}a--c. The convective granules are found to have sizes in the range 1--3\,Mm with lifetimes of the order 4--7\,min. A steady state rms convective vertical velocity obtained at the end of 120\,min is $\sim 1.54$\,km\,s$^{-1}$ at the photosphere agreeing with observations\cite{Keil78}. Beyond this, the code is run for an additional solar time of 50\,min during which the results are analysed. 
The imposed vertical magnetic field across the domain keeps the acceleration fronts more or less intact. In reality, spicules on the surface of the Sun are more prevalent in the super-granular lanes. In the 2D set-up used here, we may consider the $x$-axis to be aligned along such a super-granular lane. In the real Sun, this coherence is broken by the presence of horizontal magnetic fields (existing over longer horizontal scales ~20--30\,Mm) e.g., in inter network regions. This set-up with convection was found to generate a forest of spicules for a range of parameters, with $B_\mathrm{imp}$ varying from 12\,G--74\,G and $T_\mathrm{SL}$ in the range 0.6--1.8 million K (six independent simulation runs).

We have utilized the fully compressible higher-order finite difference numerical tool, the {\sc Pencil Code}\cite{Brandenburg02, Brandenburg20} (hosted at \verb|https://github.com/pencil-code/|) for these computations. Out of the several options available with the {\sc Pencil code}, we have chosen a sixth-order finite difference scheme and a three-step Runge-Kutta-Williamson time stepping scheme. The {\sc Pencil code} uses explicit dissipation parameters, hyper-dissipation and shock viscosity\cite{Haugen04, Chatterjee20}. 
The run directory for the 2D convective simulation\cite{Chatterjee22} contains the start (run) parameters in the files \texttt{start.in (run.in)}, 
the initial stratification file, and a reference output file. Further, the compilation file \verb|src/Makefile.local| provides the list of active physics modules, and \verb|src/cparam.local| sets the number of CPUs used as well as the grid size.  

\subsection*{Measure of anisotropy in the plasma}

We employ two separate techniques to study the effect of imposing a vertical magnetic field on the instabilities and turbulence generated in the numerical experiments. The first one is following a popular method called anisotropy invariant maps\cite{Lumley77}, where calculation of a $2\times2$ anisotropy gradient matrix is carried out at each pixel defined as 
\begin{equation}
\label{eq:aniso}
    a_{xz}=\frac{\partial_x \chi \partial_z \chi}{(\partial_x \chi)^2+(\partial_z\chi)^2}.
\end{equation}

Here, $\chi=\log T$.  However, in a general case, it can also be defined as the intensity of each pixel of an image whose anisotropy we wish to quantify.  Let the two complex eigenvalues of this matrix be $\lambda_1$ and $\lambda_2$ using which we can define two real quantities, $\xi=\lambda_1^2+\lambda_1\lambda_2+\lambda_2^2$ and $\zeta=-\lambda_1\lambda_2(\lambda_1+\lambda_2)$. In our case, the distribution (PDF) of $\xi$ over all pixels (and over all $\zeta$ values) in the domain shows an obvious trend with increasing $B_\mathrm{imp}$ when plotted in the $\zeta-\xi$ plane. The cumulative fraction of points, $f_B$, enclosed by the PDF such that the PDF$>0.1$ has been used as a measure of anisotropy. We have scaled this with the cumulative fraction of points, $f_\mathrm{ran}$, for a case when the entire image is made up of only uniformly random numbers creating a homogenous and isotropic pattern. The curve plotted in Fig.~\ref{fig:phase_plot}e is $(1-f_B)/(1-f_\mathrm{ran})$ versus $B_\mathrm{imp}/B_0$, where $B_0$ is the magnetic field at the chromospheric heights $\sim 1$\,Mm in equipartition with the kinetic energy of the harmonic driving. We perform this analysis on an image at the end of the 25\,min (solar time) since the start of the harmonic driving with a time period of 5\,min.  

The second method followed is detecting and counting of vortex patterns in the snapshots of velocity vectors employing the technique of Automated Swirl Detection Algorithm (ASDA\cite{Liu19}) using its default parameters except that the lower boundary of the peak, $|\Gamma_1|$ value, (see Eq. 1 in Ref[\cite{Liu19}]) of a vortex is defined as 0.71. This allows us to detect all vortex candidates with an average rotating speed greater than their average expanding/shrinking speed. The average number of such vortices detected in six frames sampled during a 5\,min duration are plotted.

\subsection*{Synthetic plasma emission:}

The synthetic intensity for an upper chromospheric or transition region line may be given by,
\begin{equation}
\label{eq:syn}
I_\lambda=\int G_\lambda(\rho, T) \varphi(T) dT.
\end{equation}
This expression for DEM is valid\cite{Landi08} for temperatures between $10^4$--$10^8$\,K, where, $$G_\lambda(\rho, T)=\exp\left[-\left(\log(T/T_\lambda)/w\right)^2\right]$$ is the "contribution'' function for the line with $T_\lambda=15000$\,K (comparison with Mg II k) and 80000\,K (comparison with Si IV), and $w=\log1.8$ and $\varphi(T)=(\rho/\overline{\rho})^2 ds/dT$ is the differential emission measure scaled with the square of the horizontally averaged density, $\overline{\rho}(z)$.  Similar expressions have been used\cite{Rempel16, Iijima17} for depicting synthetic coronal and chromospheric jets, respectively. Normally, the integration is performed along the line-of-sight (LOS), the infinitesimal element along which is $ds$. However, for two-dimensional runs, the integration is irrelevant as there is only one grid point along the LOS. For the visualization of the 3D run in Supplementary Video~\ref{mov:plasma_fluid}, we use the volume rendering of the quantity $dI_\lambda/ds$ for $T_\lambda=15000$\,K.
 In order to validate the simple analytical expression for $G_\lambda$, we have further made use of the FoMo code\cite{Doorsselaere16} to obtain synthetic intensity images for Mg II 279.6\,nm \,and Si IV 140.2\,nm \,lines as diagnostic tools of solar chromosphere and transition regions. This code utilizes realistic contribution function $G_\lambda(\rho,T)$ from the CHIANTI database\cite{Dere97} by considering collisional excitation and spontaneous de-excitation processes in optically thin plasmas. We would like to caution that the forward modelled Mg II intensity profiles using these methods will not be accurate because it is not exactly in the optically thin regime and the Saha-ionisation equilibrium is an approximation in that case.

Finally, the SDO/AIA 17.1\,nm emission is synthesized using the contribution function, $G_\lambda(T)$ available from the solarsoft library at \verb|https://www.lmsal.com/solarsoft/|.

\subsection*{Lagrangian Tracking}
To decipher the plasma flow inside the simulated spicules after their formation sets in, we introduce massless particles into the computational domain for Lagrangian tracking in six separate layers between $-0.5<z<6$\,Mm, each layer 1\,Mm thick (except the bottom which is 0.5\,Mm thick), represented by a different color. We observe that the particles initially within the layer $-0.5<z<1$\,Mm never rise above a height of $2$\,Mm during the 19.5 min duration of tracking, as shown in Fig.~\ref{fig:layer_track}a. However, the particles placed within the next $1<z<2$ Mm layer reach up to $6$ Mm height in the form of elongated and dense spicules.

\subsection*{Polymeric fluid experimental set-up.}

Extensive studies have been carried out by subjecting Newtonian and polymeric fluids to vertical harmonic vibrations of the container or Faraday excitation\cite{Goodridge96,Goodridge99,Wagner99} of the form $a(t)=A\sin\omega_0 t$, where $A$ is the amplitude of the external excitation. Faraday waves are parametrically driven standing surface waves whose dispersion relation is given by, $\omega^2=\tanh(kh)\left[g_\mathrm{E}k+\sigma k^3/\rho\right]$, with angular 
frequency, $\omega$, wavenumber, $k$, fluid depth, $h$, surface tension, $\sigma$ and $g_\mathrm{E}$ is the local acceleration due to gravity. The nature of the wave has a transition from surface gravity to capillary at $k'=\sqrt{g_\mathrm{E}\rho/\sigma}$ and corresponding frequency, 
$\omega'=(4g_\mathrm{E}^3\rho/\sigma)^{1/4}$, which define low-frequency gravity waves ($\omega < \omega'$) and high-frequency capillary waves ($\omega>\omega'$). For the first sub-harmonic Faraday waves, $\omega=\omega_0/2$ in the above dispersion relation. When the amplitude of excitation exceeds a certain threshold value of peak acceleration, $A_\mathrm{min}$, jet or droplet ejection occurs from the surface.

For our experiments, we use polyethylene oxide (PEO), iodinated polyvinyl alcohol (PVA) and glycerine solutions of varying viscosities. PEO solutions of 1000\,ppm concentration by weight were prepared by carefully dispersing a long-chain polymer (Alfa Aesar $>$ 5,000,000 M.W.) in powder form into deionized water to avoid clumping and stirred at low rpm along with occasional gentle shaking for three hours at room temperature ($\sim 300$\,K). Successive dilutions (20--500\,ppm) were obtained from the above master solution using deionized water and stirred similarly for about a half-hour. Polyvinyl alcohol solutions were made from commercial grade PVA (degree of hydrolysis 85--89, Thermo-Fisher Inc.). Suitable mass percentages without any further purification were slowly added 
to deionized water at the beginning while stirring continued for 3 hours at a 
temperature of 318\,K. Iodination of the solution was carried out immediately by the addition of I$_2$ (1 \,mmol l$^{-1}$) and 
KI (4\,mmol l$^{-1}$) simultaneously to the solution and stirring for a further 3 \,hr at the elevated temperature.  In both the cases, losses were made up and stirred further to achieve a single-phase solution. All the prepared solutions were allowed to sit at ambient temperature for 12 (PEO) and 24 (PVA) hr and stirred for about 15 min before the excitation, polarization, or rheometry experiments were performed. PEO solutions were used within 48\,hr, and PVA within 120\,hr, of preparation to prevent degradation.

Parallels between plasma and polymeric solutions in a different context (Magnetorotational Instability versus Elastorotational Instability in cylindrical Couette geometry) have been explored using PEO solutions previously\cite{Boldyrev09, Bai15}. The PEO solutions used here were found to have the following physical characteristics at 300\, K (concentration: $c=20, 50$ and $100$\,ppm, density: $995$ kg\,m$^{-3}$, kinematic viscosity: $\nu=9.8\pm 0.1\times10^{-7}$\,m$^2$s$^{-1}$ and for $c=500$\,ppm, $\nu=2.0\pm 0.1\times10^{-6}$\,m$^2$s$^{-1}$ whereas for $c=1000$\,ppm, $\nu=5.5\pm0.2\times10^{-6}$\,m$^2$s$^{-1}$ ). The viscosity measurements were carried out using an MCR 302 rheometer from Anton Paar with shear rates varying between 1--200\,Hz. The concentration, $c^*$, at which the chain-chain interaction in PEO solution cannot be neglected occurs when the solution viscosity is almost twice the solvent viscosity, is about 500\,ppm. The physical parameters for Iodinated 2\% PVA solution were found be
(density: $995$ kg\,m$^{-3}$, kinematic viscosity: $\nu=3.2\pm0.2\times 10^{-5}$\,m$^2$s$^{-1}$ and, surface tension, $\sigma$: $0.052$ N m$^{-1}$). Iodinated 3\% PVA solution yielded similar values of density ($\rho=982$ kg\,m$^{-3}$) and $\sigma=0.053$ N m$^{-1}$ but $\nu=5.7\pm0.2\times 10^{-5}$\,m$^2$s$^{-1}$. A 76\% solution of glycerine (99.5\% Sigma Aldrich) was prepared by diluting with appropriate amount of deionized water and stirred at 300\,K for 20 min ($\rho$: 1195\,kg\,m$^{-3}$, viscosity: $4.0\pm0.2\times 10^{-5}$\,m$^2$s$^{-1}$, surface tension: 0.064\,N\,m$^{-1}$). 

Measurements of jet heights were performed during an interval of ten excitation periods after the ejection of the geometric central jet and prior to bubble formation. Observation of the forest of jets and related nonlinear effects are robust over a range of geometry, types of fluid, viscosity, and forcing amplitude investigated by us. 

\subsection*{Excitation experiments}

The excitation experiments were performed on a subwoofer speaker (12 inch 4 Ohms 200 W R1S4-12 Rockford Fosgate), powered by a 100 W audio bass amplifier. The power fed to the speaker never exceeded 60 W (fused) to allow for a linear electrical response, also to allow it to act as a Faraday shaker. Geometrical effects were not found to be significant as similar behaviour was observed on a smaller speaker with identical electrical characteristics. A levelled shallow cylindrical vessel of 0.1\,m diameter was firmly anchored to the centre of the driver cap of the speaker and a fluid volume of 30\, ml was added. Provision was made to measure the acceleration using a calibrated MPU 9250 (TDK Invensense) accelerometer as well as from high speed imaging with or without loading. All the experiments have been performed at $300 \pm 1$\,K. Our peak accelerations are in the range $3-15 g_\mathrm{E}$ and we find a sub-harmonic response of the surface in all cases, even though the pattern changes over the number of periods depend on the forcing.

\subsection*{Polymer stretching on excitation}

The vertical stretching of polymer chains under large excitation was observed using the same set-up. The region of interest above the vessel was shone using an enlarged cross-section He-Ne laser beam (632.8\,nm) with the region of interest placed between two crossed polarizers.
We use two different configurations of the crossed polarizers (Extended Data Fig. \ref{fig:pol_setup}a): i) 0$^\circ$-90$^\circ$ (dark mode) in which the axis of the polarizer is aligned with the vertically polarised source and that of the analyzer is perpendicular, ii) 45$^\circ$-135$^\circ$ (bright mode) where the polarizer-analyzer configuration is rotated by $45^\circ$ to the vertical. Because of absorption by Iodine in the stretched polymer chains (Extended Data Fig. \ref{fig:pol_setup}b), the beam suffers extinction in the first configuration, thus resulting in a dark field both in the presence or absence of the jets. On the other hand, when the polarizer axis is turned by $45^\circ$, with the analyzer turned  $135^\circ$ (Extended Data Fig. \ref{fig:pol_setup}c), illumination is observed only in the presence of jets that offer a polarizing medium at $45^\circ$ to both the polarizer and the analyzer. 

The various non-dimensional numbers related to jetting in our experiments are as follows. The Ohnesorge number, $Oh=\nu\sqrt{\rho/\sigma d}$, where $d$ is the typical jet diameter, comparing the visco-capilliary time scale, $t_v=\rho \nu d/\sigma$, to Rayleigh time scale for inviscid capillary break-up, $t_c=(\rho d^3/\sigma)^{1/2}$, takes the following values: 0.003 (water/PEO 10--100\,ppm, surface tension: $\sigma \approx 0.070$ N m$^{-1}$), 0.005 (PEO 500\,ppm), 0.098 (Iodinated PVA) and 0.12 (glycerine solutions) at 30\,Hz and a forcing acceleration of $\sim 10g_\mathrm{E}$. The Weber number, $We=\rho U^2 d/\sigma$, which is a measure of inertial forces to surface tension in multi-phase fluids, is in the range 4--6.  In the frequency and forcing amplitude ranges covered in this study, we do not see any formation of neck or jet pinching in excitation experiments with PEO solutions of $c> 50$\,ppm, in spite of $Oh<<1$, in contrast to water (similar low viscosity). The typical strain rate experienced by polymer chains, during the rise time of the jets ($\sim 16$\,ms), is given by, $U_\mathrm{jet}/L_\mathrm{jet}\sim 13.3$\,s$^{-1}$, where $U_\mathrm{jet}$ and $L_\mathrm{jet}$ are average velocity and maximum height reached by a long jet, respectively. In 2\% PVA solution versus 76\% glycerine solution experiments (similar but higher viscosity than PEO solutions), we observe neck formation and pinching. We note less jet breakage in PVA (23\% PVA jets reaching heights above 1\,cm break as compared to 53\% in glycerine solution) indicating that uncoiling polymer chains may be suppressing Plateau-Rayleigh instability. 
\subsection*{Observational data}

We used a series of spicule image sequences obtained with the IRIS (\href{https://iris.lmsal.com/}{Interface Region Imaging Spectrograph}) at 11:24:46, on February 21, 2014\cite{Pereira14}. The investigated area is located in the south pole of the Sun with the center solar coordinates of  $x_S=7$\arcsec and $y_S=969$\arcsec  with a field-of-view (FOV) of 119\arcsec$\times$119\arcsec.
For our work, we analysed spicules observed with the transition region Si IV 140.0\,nm (at T$\approx$80000\,K) and the chromospheric Mg II k 279.6\,nm (at T$\approx$15000\,K)  slit-jaw filtergrams\cite{Skogsrud15} with a 19 s cadence. The data extracted is for a duration of 20 min from 11:24:46 UT.

In Fig.~\ref{fig:sun_lab}a (for Si IV) and Extended Data Fig.~\ref{fig:MgII}b (for Mg II k), with the images in $[x,z]$--plane, we focus on an area of $80\times20$\,Mm$^{2}$ from the slit-jaw filtergram data that were taken at 11:34:16, on February 21, 2014. The blue rectangle highlights an area with the same extent as the simulation box. To reveal the bright tip of some of the spicules, first, we determine the horizontally averaged intensity, $I_0$, for each pixel in the $z$-direction. 
Next, we construct an intensity map scaled by $I_0$ for each pixel\cite{DePontieu07}.
To visualise the evolution of a spicule (denoted by a white arrow, S1) in the Si IV and Mg II k images of Fig.~\ref{fig:sun_lab} and Extended Data Fig.~\ref{fig:MgII}, respectively, time-distance plots were constructed. 
\section*{Data Availability}
The data points for Fig.~\ref{fig:phase_plot}a,b,e and f and histogram data for Extended Data Fig. 6 are provided as ascii files. The data for Fig.~\ref{fig:sun_lab}a and Extended Fig.~\ref{fig:MgII}a,b can be downloaded from \href{https://iris.lmsal.com/}{iris.lmsal.com}. The raw simulation data files in Fortran binary format (about 300 GB in size) used to produce Fig.~\ref{fig:sun_lab}b-c is available from the corresponding author upon reasonable request.

\section*{Code Availability}
The Pencil code is hosted at \verb|https://github.com/pencil-code/|. The run directory for simulation with solar convection represented in Fig.~\ref{fig:sun_lab}b is publicly available at DOI: \href{https://doi.org/10.5281/zenodo.5807020}{10.5281/zenodo.5807020} upon request.

\newpage

\clearpage
\renewcommand{\baselinestretch}{1.}
\clearpage

\begin{table}[h!]
\renewcommand{\tablename}{Extended Data Table}%
  \centering
  \caption{\label{tab:tab1} {\bf Dynamical similarity:} List of non-dimensional parameters with respect to which the two systems can be considered dynamically similar. Here, $d$ and $L$ are the typical width and length of the jets, whereas, $U$ is the typical velocity scale. The Froude number indicates the importance of surface gravity in the dynamics. The expression for Froude number is different for two different scenarios - one for a continuum stratified solar atmosphere and the other for a fluid-air interface (polymeric fluid). The density scale height in the solar chromosphere is denoted, $H_\rho \sim 150$\,km. The peak acceleration of the Faraday excitation (fluid) and chromospheric plasma acceleration, in terms of the local gravity, is given by $a/g$. Here, $g$ takes value $g_\mathrm{S}\sim 274$\,m\,s$^{-2}$ for the Sun and $g_\mathrm{E}\sim 9.8$\,m\,s$^{-2}$ for the Earth. In the solar plasma, the frequency of driving is $f_0=3.3$\,mHz whereas, in the fluid, $f_0=30$\,Hz and $\tau=L/U$ is the typical lifetime of the jets. The fluids are 2\% and 3\% PVA solution.}
  \begin{tabular}{@{}llllrrrr@{}}
  	\\
  	\toprule
  	Jet & $d$ & $L$ & $U$ & Aspect ratio & Froude number & $a/g$ & $f_0\tau$ \\
  	& & & & $d/L$ & & & \\
  	\midrule
  	Plasma& 0.5--1\,Mm & 7--20\,Mm & 30--50\,km\,s$^{-1}$ & 0.05--0.15 & $U\sqrt{\frac{H_\rho}{g_\mathrm{S} d^2}}$& 4--20 & 1.0--1.8 \\
  	& & & & & $\sim$ 0.7--2.3 & \\
  	Fluid & 0.2--0.42\,cm& 1.5--4.0\,cm& 0.25--0.4\,m\,s$^{-1}$ & 0.07--0.21 & $U/\sqrt{g_E d}$  & 3--15 &2.0--3.0 \\
  	& & & & &  $\sim$ 1.2--2.2 \\
  	\bottomrule
  \end{tabular}
\end{table}
\begin{table}[h!]
\renewcommand{\tablename}{Extended Data Table}%
  \centering
  \caption{\label{tab:tab2} {\bf List of runs:} Parameters of the numerical and the table top runs along with the figures and videos where they have been referenced. All the solar MHD simulations are performed with a sponge layer temperature, $T_\mathrm{SL}=10^6$\,K. The driving velocity amplitude is $V_0$ at the driving frequency, $f_0$. The wave number $k=n\pi/L$, where $n$ is the number of crests in the horizontal domain of length, $L$. For the polymeric fluid experiments, $A_0$ denotes the peak acceleration amplitude scaled with local gravity, $g_\mathrm{E}=9.8$\,m\,s$^{-2}$, $c$ is the polymer concentration expressed either as parts-per-million (ppm) or as weight percentage, $\nu_p$ is the measured kinematic viscosity.} 
  \begin{tabular}{@{}rllllll@{}}
  \\
    \toprule
    & {\bf Solar plasma},&$T_\mathrm{SL}=10^6$\,K &  & & & \\
    \toprule
 {\em Dimension}  &{\em Forcing} & $V_0$ (\,km\,s$^{-1}$) & $f_0$(\,mHz) & $n$ & $B_\mathrm{imp}$\,(G) & {\em Reference}  \\
    \midrule
\ldelim \{{8}{*}[2D] & wave driving &0.45--2.50 & 3.0--9.0 & 5.0 & 12 & Fig.~\ref{fig:phase_plot}a\\
  & wave driving & 1.32& 3.3 & 5.0 & 0--1 & Fig.~\ref{fig:phase_plot}c,e\\  
  & wave driving & 1.0 & mixed & 5.0 & 12 & Video~\ref{mov:mixed}\\
  & convection & & & & 74 & Fig.~\ref{fig:sun_lab}b--c, ~\ref{fig:layer_track}a--d,  \\
  & & & & & & Ext. Figs.~\ref{fig:MgII}c--d,
  \\
   & & & & & & \ref{fig:Ma_shock},~\ref{fig:spicheat}a--c,~\ref{fig:hist}a,~\ref{fig:azvz_track_Lorentz}, ~\ref{fig:convection},\\
  & & & & & & Videos~\ref{mov:SiIV},~\ref{mov:tracking}\\
  &  wave driving & 1.32 &3.3 &5.0 & 0 &  Ext. Fig.~\ref{fig:zero_mag}a--b\\
 3D & wave driving  & 1.32 & 3.3& 1.0 & 10 & Video~\ref{mov:plasma_fluid}\\
 \toprule
     & {\bf Polymeric fluid} & & & & & \\
    \toprule
{\em Type} & {\em Solution}  &  $A_0/g_\mathrm{E}$ &  $f_0$ (Hz)& $c$ (ppm) & $\nu_p$ ($10^{-7}$\,m\,$^{2}$\,s\,$^{-1}$) & {\em Reference}  \\ 
 \toprule
\ldelim \{{8}{*}[Polymeric]   &PEO &1.0--10.1& 15--120 &50, 500 & 9.8$\pm$0.1,20.0$\pm$1.0 & Fig.~\ref{fig:phase_plot}b\\
  &PEO &10.0&30 & 10-100 & $9.8\pm0.1$ &Fig.~\ref{fig:phase_plot}f\\
  &PEO &4.9,13.2&30,120 & 1000 & $55.0\pm2.0$ & Video~\ref{mov:jetbreaking}a--b \\
 
 &  Iod. 2\% PVA & 10.1& 30&2\% wt.& 320.0$\pm$20.0 &Ext. Fig.~\ref{fig:pol_setup},~\ref{fig:hist}b--c \\
 & & & & & & Video~\ref{mov:jetbreaking}c\\
 &  Iod. 2\% PVA & 5.0 & mixed &2\% wt.& 320.0$\pm$20.0 &Video~\ref{mov:pmode}\\
 &  Iod. 3\% PVA & 10.1 & 30 &3\% wt & 570.0$\pm$ 20.0 &  Video~\ref{mov:plasma_fluid},~\ref{mov:mechanism}\\
 & Acrylic Paint & 7.0 & 30 & & & Fig.~\ref{fig:sun_lab}d, Video~\ref{mov:paint}\\
\ldelim \{{3}{*}[Newtonian]  & Water & 10.0 & 30 &0& 9.8$\pm$0.1 &Fig.~\ref{fig:phase_plot}d\\
 &  Glycerine &1.0--10.1 & 15--120 & 55\% wt. & 55.0$\pm$2.0 & Fig.~\ref{fig:phase_plot}b,
 Video~\ref{mov:jetbreaking} \\
 &  Glycerine & 10.1 & 30 & 76 \% wt. & 400.0$\pm$20.0 & Video~\ref{mov:jetbreaking}\\
   \bottomrule
  \end{tabular}
\end{table}

\clearpage

\setcounter{figure}{0}  
\begin{figure}
\renewcommand{\figurename}{Extended Data Figure}%
\includegraphics[width=0.50\textwidth]{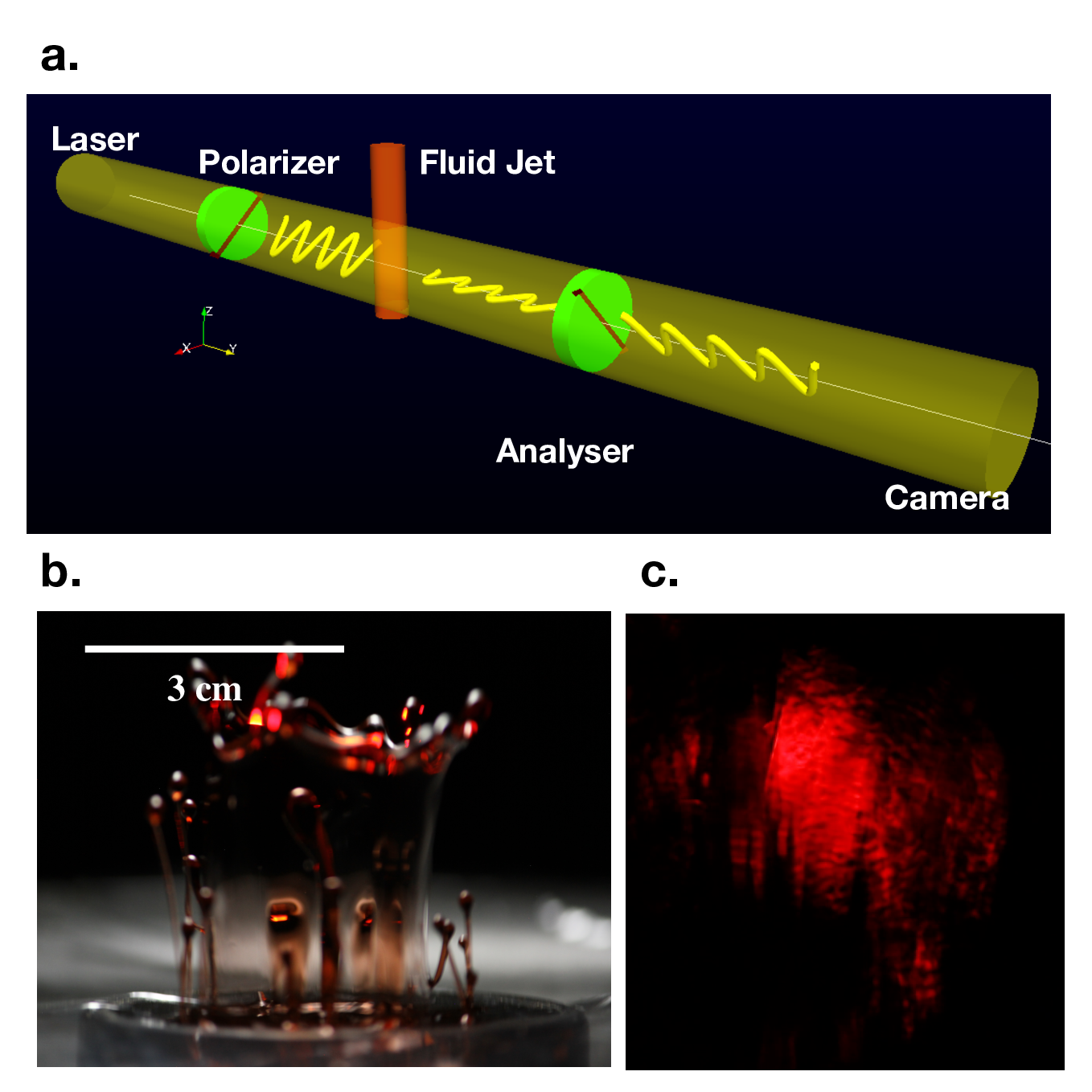}
\caption{\label{fig:pol_setup} {\bf Signature of polymer stretching} (a) The set up for the polarization experiment for detecting polymeric stretching in 2\% Poly Vinyl Alcohol. He-Ne laser light passes through a polarizer and then through the sample region of interest probing the jets, to be collected after an analyzer by an imaging device. By varying the angle of the polarizer-analyzer combination at right angles to each other, the polarization state of the jets can be determined (see Methods). (b) Jets of iodinated 2\% aqueous PVA by Faraday excitation at 30 Hz. (c) Similar to (b) but seen through the polarizer set-up in (a) to highlight stretching of polymer threads.}
\end{figure}
\begin{figure}
\includegraphics[width=0.50\textwidth]{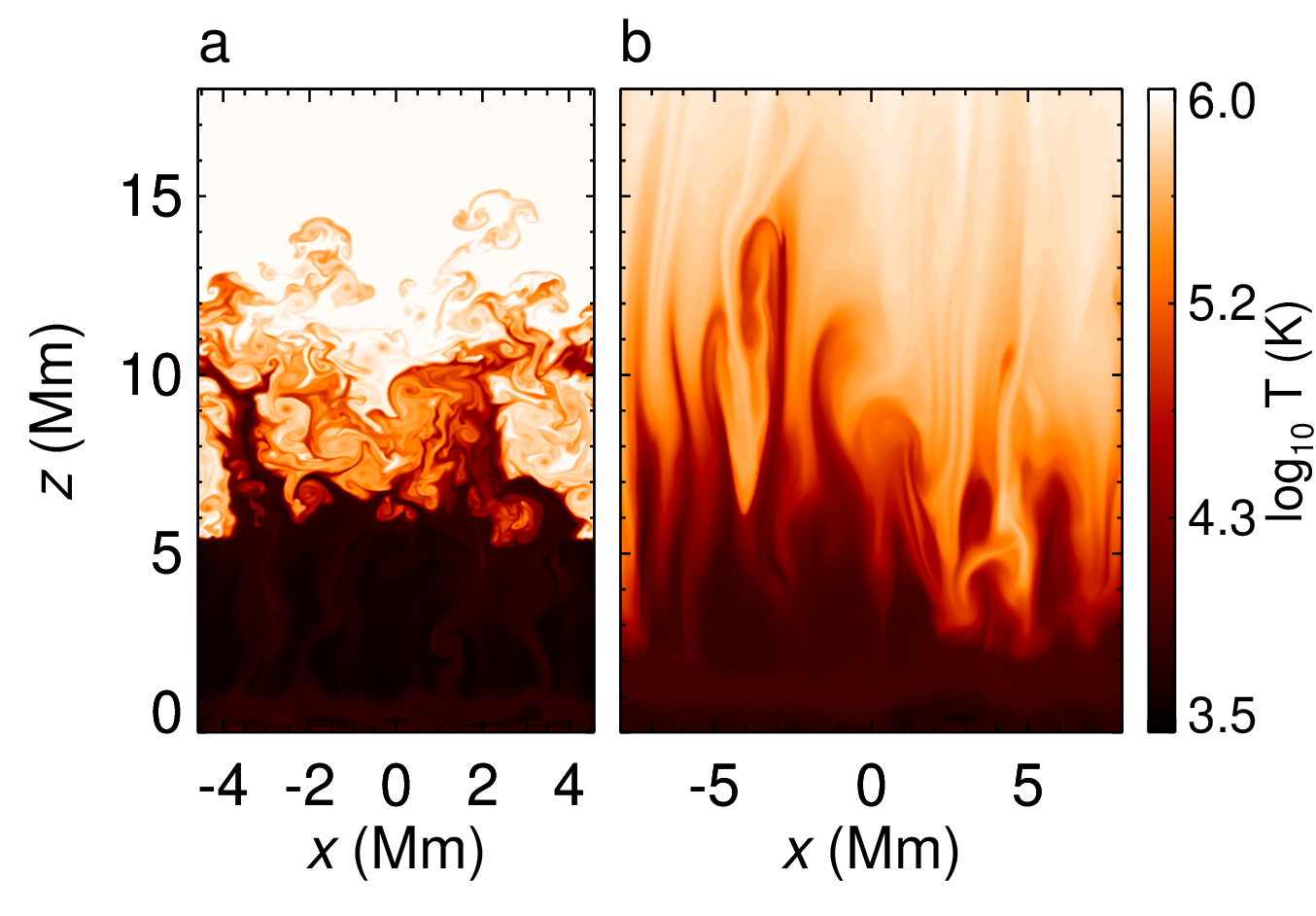}	
\renewcommand{\figurename}{Extended Data Figure}%
\caption{\label{fig:zero_mag} {\bf Kelvin-Helmholtz instability and magnetic field} (a) Contours of temperature in the simulation domain with zero imposed magnetic field shows turbulence generated by the Kelvin-Helmholtz instability. (b) Same as (a) with the addition of a damping term in the horizontal velocity equation to mimic suppression of turbulence in the horizontal direction.}
\end{figure}

\begin{figure}
\renewcommand{\figurename}{Extended Data Figure}%
\includegraphics[width=0.85\textwidth]{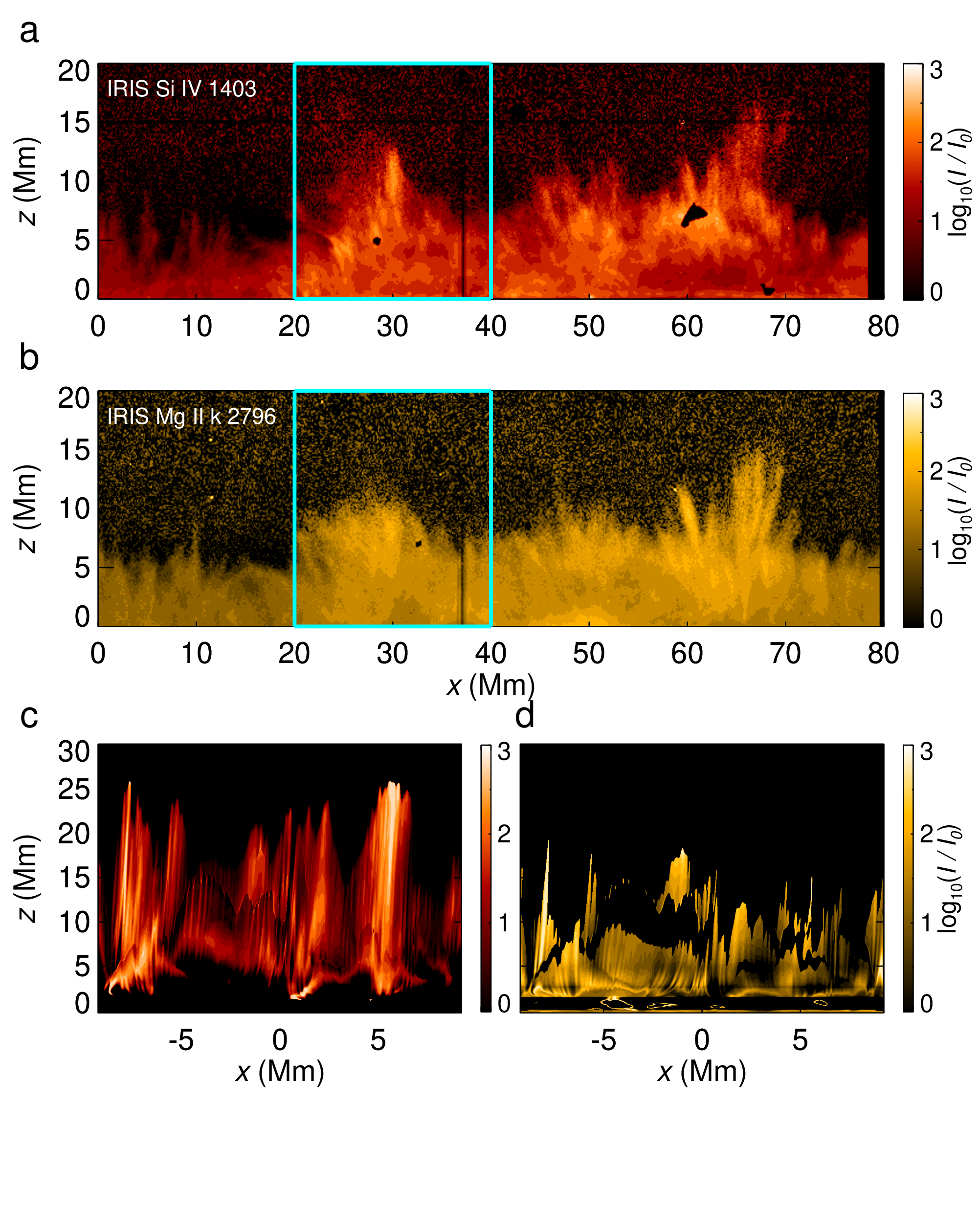}	
\caption{\label{fig:MgII} {\bf Observed and simulated spicules seen in Si IV and Mg II emission lines} (a) Spicules seen at the solar limb with the Si IV filter of IRIS. (b) Same as (a) but for IRIS Mg II k filter. (c) Shaded contours of synthetic intensity of Si IV emission synthesized using the CHIANTI package available in the FoMo code, for the simulation run with imposed vertical magnetic field $B_\mathrm{imp}=74$ G at time, $t=152.50$ min from the start are shown in the $[x, z]$-plane. (d) Same as (c) but for emission at Mg II k also using CHIANTI. The cyan rectangle in panels (a) and (b) indicates the same extent as the simulation domain shown in (c)-(d).}
\end{figure}

\begin{figure}
\renewcommand{\figurename}{Extended Data Figure}%
\includegraphics[width=0.60\textwidth]{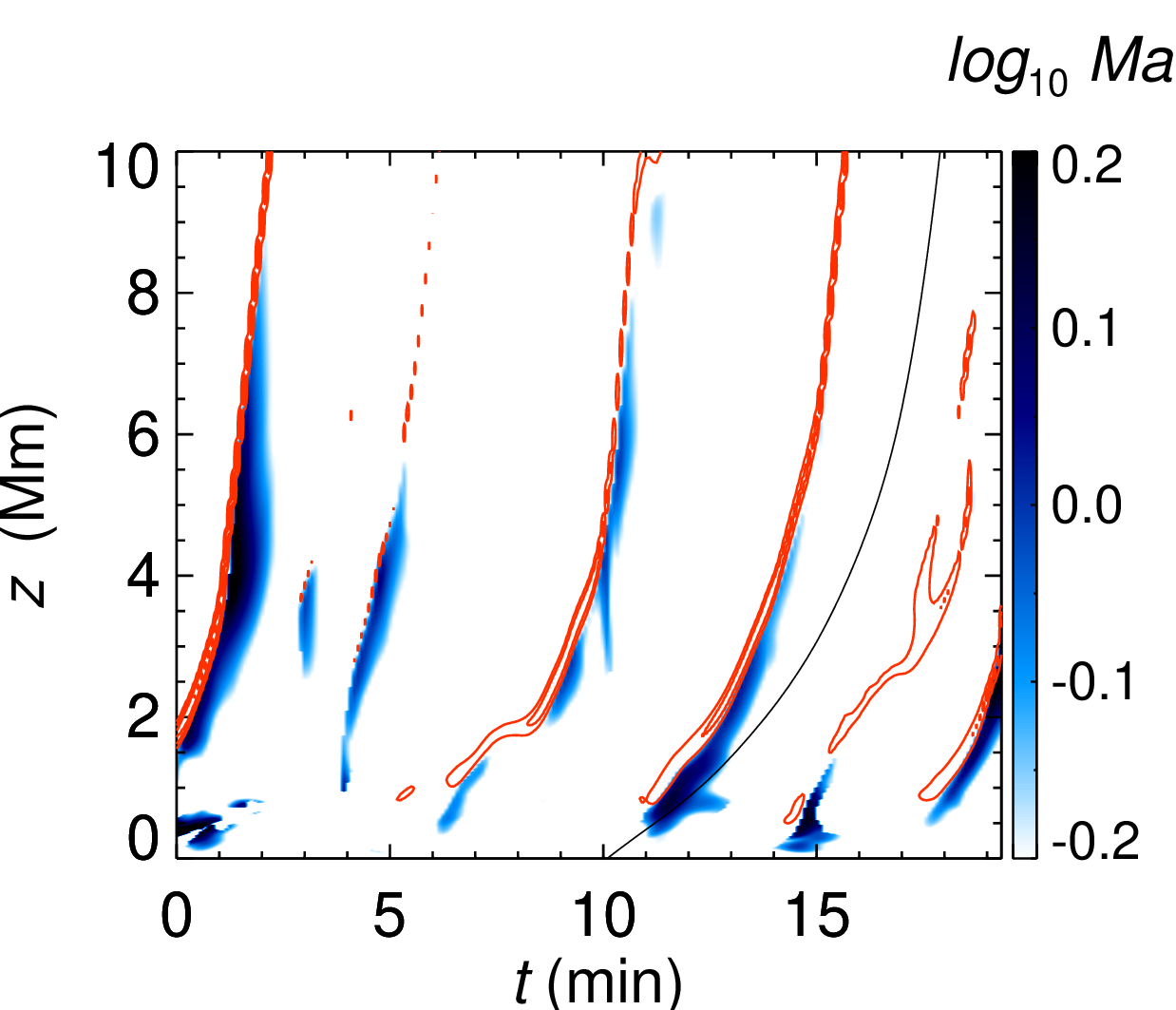}	
\caption{\label{fig:Ma_shock} {\bf Propagation of shock compression and hyper-sonic regions} (a) Time-distance map of the Mach number, $Ma$ (blue shaded), and shock compression (red contour lines) for a vertical slit at $x=2.5$\,Mm of panel (a). The regions of plasma compression, $\nabla\cdot\mathbf{V}<0$, are found to coincide with acceleration fronts, shown in Fig.~\ref{fig:layer_track}c, where the spicular plasma is energised to hypersonic speeds. The characteristic curve for acoustic waves is denoted by the solid black line.}
\end{figure}

\begin{figure}
\renewcommand{\figurename}{Extended Data Figure}%
\includegraphics[width=0.55\textwidth]{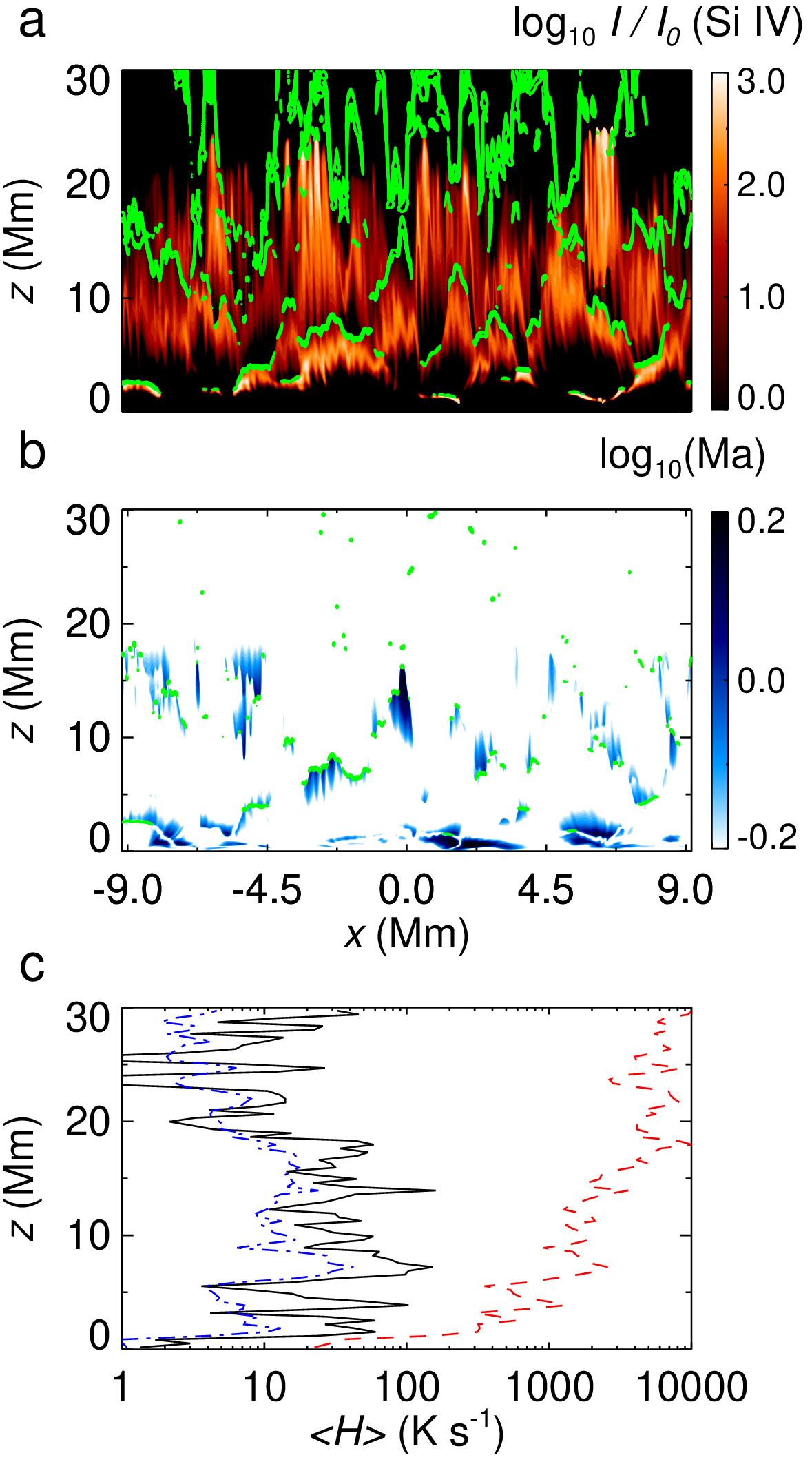}	
\caption{\label{fig:spicheat} {\bf Plasma heating processes} (a) Synthetic Si IV emission from the simulated spicule forest similar to Fig.~\ref{fig:MgII}c calculated using CHIANTI. The green contours denote the locations of plasma heating above $10^3$\,K\,s$^{-1}$ only due to the local compression of plasma, $-(\gamma-1)T\nabla\cdot{\mathbf V}$, where $T$ and $\mathbf V$ are the plasma temperature and velocity, respectively. The ratio of specific heats denoted, $\gamma=c_P/c_V$, has been calculated using the equation of state for a partially ionized ideal gas. (b) Regions of Mach number, $Ma >1$, are indicated by shaded blue contours. Here, the green contours show heating above $10^3$\,K\,s$^{-1}$ only due to dissipation inside the shocks by $\zeta_\mathrm{shock}(\nabla\cdot{\mathbf{V}})^2/c_V$, where $\zeta_\mathrm{shock}$ is the shock viscosity. (c) Strength of various heating sources as function of height in the atmosphere. Horizontally averaged heating profiles, $\langle H\rangle(z)$, due to local plasma compression (red-dashed), shock dissipation (black-solid), and Ohmic heating (blue dot-dashed).}
\end{figure}

\begin{figure}
\renewcommand{\figurename}{Extended Data Figure}%
\includegraphics[width=0.40\textwidth]{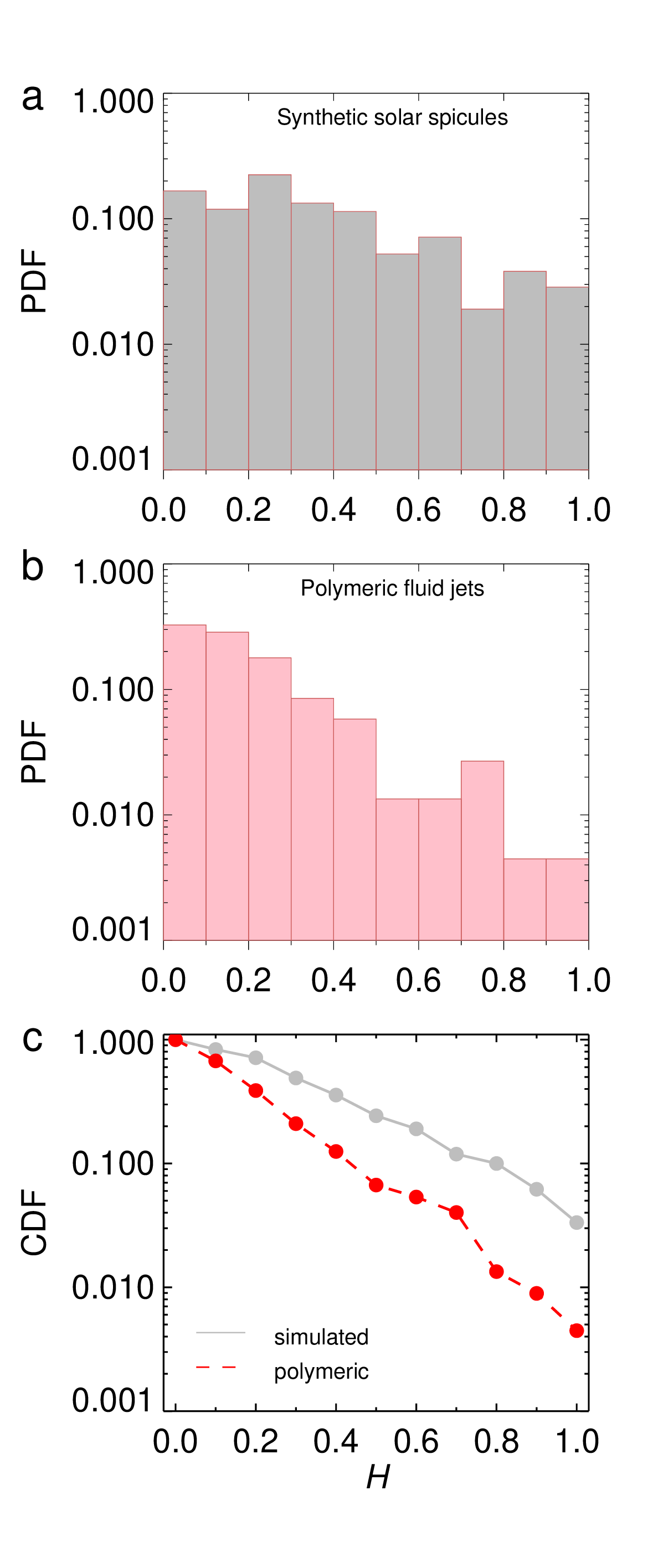}	
\caption{\label{fig:hist} {\bf Comparing height distribution of solar spicules and fluid jets} (a) Histogram for distribution of $H=(h-h_\mathrm{min})/(h_\mathrm{max}-h_\mathrm{min})$, where $h$ is the maximum height reached by a spicule and $h_\mathrm{min}$, $h_\mathrm{max}$ are the minimum and maximum heights of a sample of unique spicules (sample size 210 and counted during 40\,min of solar time), respectively, as seen in synthetic emission at $T=15,000$\, K from the simulation shown in Extended Fig.~\ref{fig:MgII}b. (b) The distribution of $H$ from a sample of unique fluid jets (sample size 224 and counted during 10 excitation periods post the central jet ejection) in an experiment with iodinated 2\% PVA. Note that only spicules such that $6 < h < 24.9$\, Mm and fluid jets such that $0.5 < h < 4.5$\,cm have been included in the distribution. The lower limit in the simulation is chosen because individual spicules cannot be discerned if they are below 6\,Mm, whereas, in the experiment the lower end is set by the container height of 0.5\,cm. The upper limits are the maximum heights obtained in the experiments.(c) The cumulative distribution function (CDF) to indicate the number of jets above $h>h_\mathrm{min}$ for both plasma and polymeric fluid. }
\end{figure}

\begin{figure}
\renewcommand{\figurename}{Extended Data Figure}%
\includegraphics[width=0.70\textwidth]{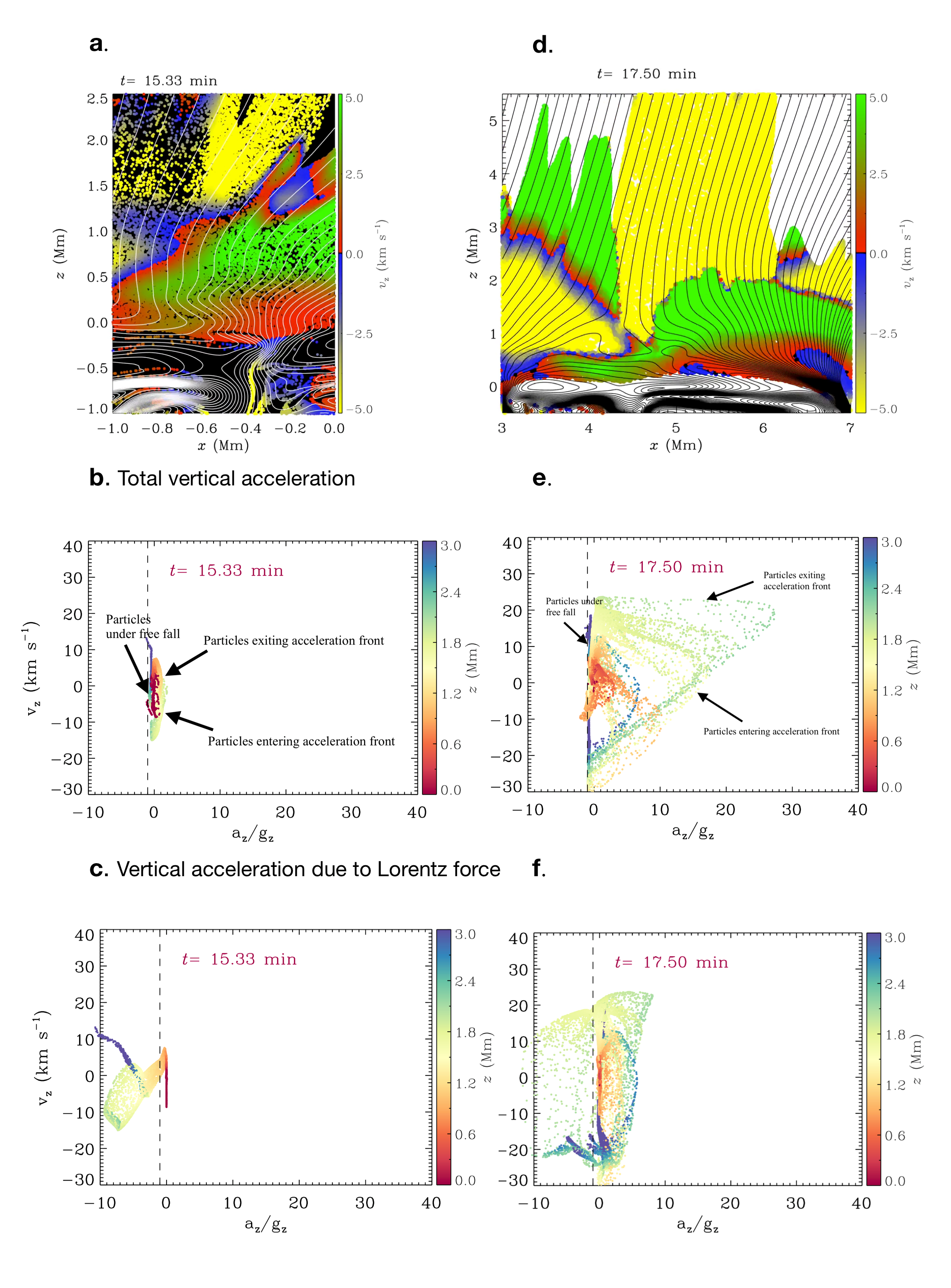}	
\caption{\label{fig:azvz_track_Lorentz} {\bf Phase plot of vertical component of velocity, $v_z$ versus  acceleration, $a_z$.} (a) The region around a shorter spicule above an inter granular lane squeezing site used for the calculation at $t=15.33$\,min from the start of tracking. Each dot denotes a Lagrangian particle colored by its instantaneous vertical velocity, $v_z$. The yellow contours are magnetic field lines. (b) $v_z$ versus $a_z$. (c) $v_z$ versus the vertical acceleration only due to the Lorentz force. Note that the particles considered, initially at $t=0$ lie in layers $0<z<3$\,Mm. Every dots denotes a particle and colored by its instantaneous vertical position. (d) The region where a longer spicule takes birth just above a granule at $t=17.50$\,min. The magnetic field lines are shown by black contours. (e) and (f) are same as (b) and (c), respectively, but, for the case of the longer spicule shown in (d).}
\end{figure}

\begin{figure}
\renewcommand{\figurename}{Extended Data Figure}%
\includegraphics[width=0.55\textwidth]{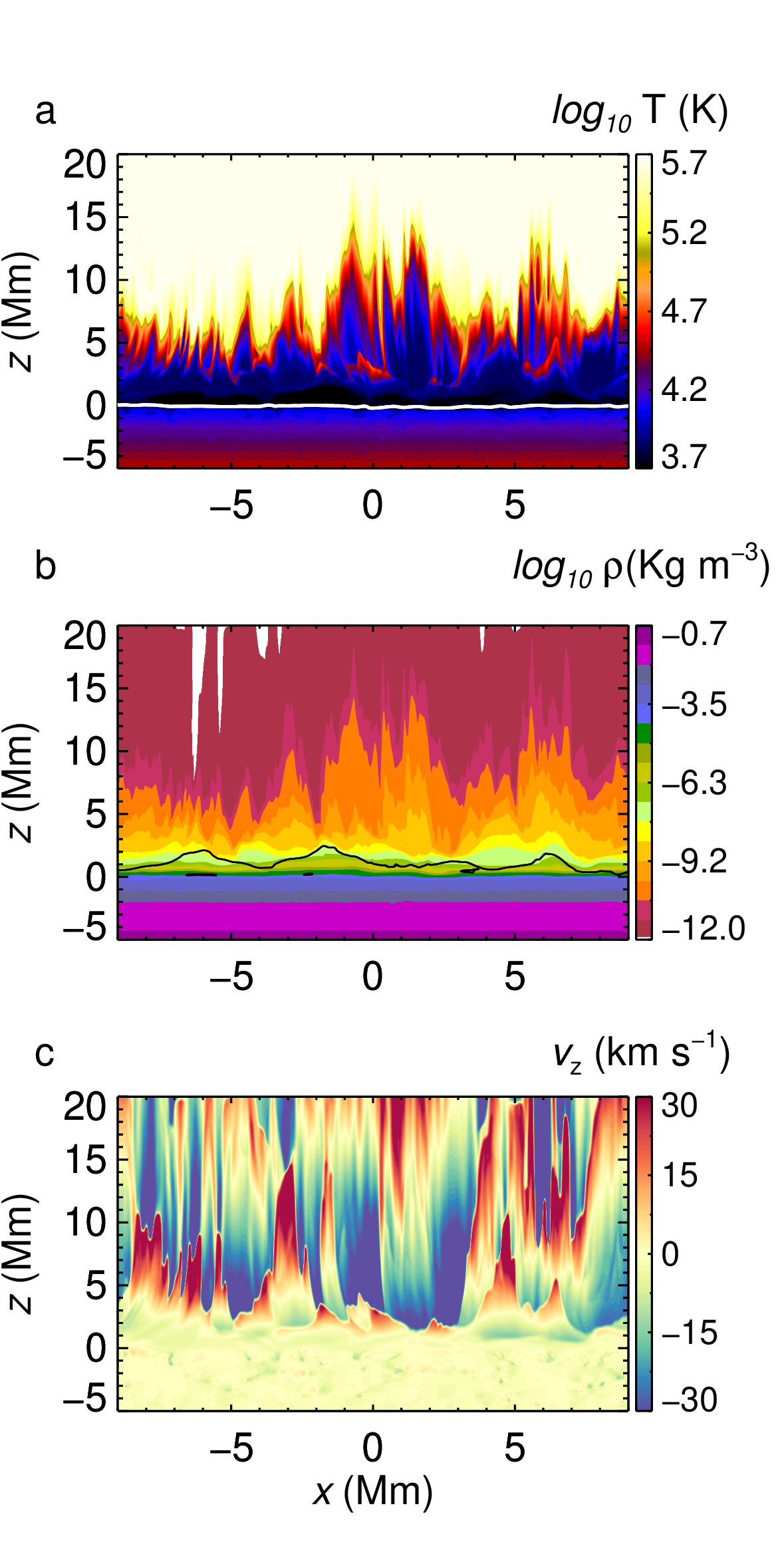}	
\caption{\label{fig:convection} {\bf Solar atmospheric stratification with convection} (a) Temperature structure at the time of introduction of Lagrangian particles in the domain for the same run shown in Fig.~\ref{fig:layer_track}a. The white solid line is the optical depth, $\tau=0.1$ surface. (b) The density at the same instant as (a). The plasma $\beta=1$ surface is denoted by the black solid line. (c) The vertical velocity, $v_z$ corresponding to panels (a) and (b). }
\end{figure}

\setcounter{figure}{0}

\begin{figure}
\renewcommand{\figurename}{Video}%
\renewcommand{\thefigure}{\arabic{figure}}
    \caption{\label{mov:mixed} {\bf Response of solar plasma to superposition of frequencies}: (a) Synthetic Si-IV emission according to the analytical expression given by Eq.~\ref{eq:syn} from a model driven by a velocity forcing of the form $v_z/V_0=\sin(2\pi f_1 t)/4+\sin(2\pi f_2 t)/2+\sin(2\pi f_3 t)/4$ shown in panel (c), where $V_0=1.0$\,km\,s$^{-1}$, $f_1=3.0$\,mHz, $f_2=3.5$\,mHz, and $f_3=4.0$\,mHz. The above forcing has only been applied within a layer, $-50$\,km $<z<0$\,km. (b) Same as (a) but for temperature (in Kelvin).}
\end{figure}

\begin{figure}
\renewcommand{\figurename}{Video}%
    \caption{\label{mov:jetbreaking} 
    Jet breaking in polymeric fluid (PEO, 1000ppm, viscoelastic) versus 55\% glycerine solution (viscous) at two different frequencies (a) 120\,Hz, and (b) 30\,Hz with acceleration amplitudes in terms of local gravity indicated. (c) Polymeric fluids (iodinated 2\% PVA, 23\% jets reaching heights above 1 cm break) may resist droplet ejection by pinching of jets (due to Plateau-Rayleigh instability) as compared to Newtonian fluids (76\% Glycerine solution, 53\% jet-breakage above 1 cm) of similar viscosity and surface tension. Both fluids are subjected to similar peak forcing of $10.1g_\mathrm{E}$ (30\,ml fluid in cylindrical container).}
\end{figure}

\begin{figure}
\renewcommand{\figurename}{Video}%
    \caption{\label{mov:plasma_fluid} {\bf Nonlinear development in numerical and table-top experiment}: Volume rendering of synthetic emission at 15000\,K for a three-dimensional numerical simulation of solar plasma forced at the lower boundary with a periodic forcing with a velocity amplitude of 1.32\,km s$^{-1}$ and an imposed magnetic field of 10\,G (left panel).
    A polymeric fluid (iodinated 3\% PVA solution) subjected to Faraday excitation at 30 Hz and a peak acceleration of $10.1g_\mathrm{E}$ (right panel). Both the numerical and the table-top experiments show nonlinear development in about five oscillation periods. The rectangular simulation domain has been clipped to have a circular cross-section using the visualization software, Paraview, to match the experimental setup.}
\end{figure}

\begin{figure}
\renewcommand{\figurename}{Video}%
    \caption{\label{mov:mechanism} {\bf Mechanism of formation of fluid jets}: View of the free surface of iodinated 3\% PVA, as seen at an angle of $20^\circ$ from the vertical, to show the mechanism for jet forest formation by collision of ridges of neighbouring polygonal cells formed due to interacting surface gravity waves (via nonlinear mode coupling).}
\end{figure}

\begin{figure}
\renewcommand{\figurename}{Video}%
\renewcommand{\thefigure}{\arabic{figure}}
    \caption{\label{mov:pmode} {\bf Response of polymeric fluid to frequency scaled solar p-modes}: Iodinated 2\% PVA (30 ml) placed in a circular container of 0.1\,m diameter and responding to frequency-scaled quasi-periodic solar acoustic modes (p-modes from SOHO/MDI) such that the dominant FFT peak lies at 30 Hz. The peak acceleration reaches up to $5 g_\mathrm{E}$. Imaged at 500 fps.}
\end{figure}

\begin{figure}
\renewcommand{\figurename}{Video}%
    \caption{\label{mov:SiIV} Synthetic emission at a temperature of 80000\,K calculated using the analytical expression of Eq.~\ref{eq:syn} (top panel) and AIA 17.1\,nm emission from our numerical simulation (bottom panel). Magnetic fields are shown by red curves. }
\end{figure}

\begin{figure}
\renewcommand{\figurename}{Video}%
    \caption{\label{mov:paint} {\bf Faraday excitation experiment with paint}: Initially separate red (15\,ml) and white (15\,ml) paint in a circular container of diameter 0.1\,m  responding to a 30\,Hz harmonic Faraday excitation with a peak acceleration of 7$g_\mathrm{E}$ imaged at 1000 fps. We observe the abundant formation of jet forest, some of which exhibit rotation as visualized by twisted motion of the red and white threads of paint.
    }
\end{figure}

\begin{figure}
\renewcommand{\figurename}{Video}%
    \caption{\label{mov:tracking} {\bf Lagrangian tracking of the plasma in the spicule forest}: (a) 540000 particles are initially distributed uniformly in six differently coloured layers, each with 1\,Mm thickness except the bottom one which is 0.5\,Mm thick, between $-0.5<z<5$\,Mm. (b) and (c) Particles are tracked by local acceleration and vertical velocity to show acceleration fronts energizing the spicular plasma. Black curves in (b) show magnetic field lines whereas the black contours in (c) denote regions of shock. (d) Spicule formation driven by granular collapse in convective plumes, P1 and P2. (e) Spicule formation driven by solar global oscillations and magnetic reconnection (denoted by the white rectangle).}
\end{figure}

\begin{figure}
\renewcommand{\figurename}{Video}%
    \caption{\label{mov:threshold} {\bf Determination of $A_\mathrm{min}$ in $A_0$ versus $f_0$ phase space}: Part I (solar atmosphere): Spicules formed in response to wave driving  of the form $2\pi f_0 V_0\sin(2\pi f_0 t)\cos(5\pi x/L)$, with (a) $V_0=0.5$\,km\,s$^{-1}$, (b) $V_0=0.35$ and (c) 0.25\,km\,s$^{-1}$, respectively. The dashed line indicates the 7\,Mm height. Note that the forest of jet criteria is satisfied for cases (a) and (b) only. Part II (Polymeric fluid): Jets formed in response to Faraday excitation with peak accelerations, $A_0$ as (a) $4.1 g_\mathrm{E}$, (b) $3.6 g_\mathrm{E}$, (c) $4.3 g_\mathrm{E}$, and (d) $4.7 g_\mathrm{E}$. The dotted line denotes the 1.3\,cm mark on the scale. Only (c) and (d) satisfy the forest of jets criteria.}
\end{figure}
\includepdf[pages=-]{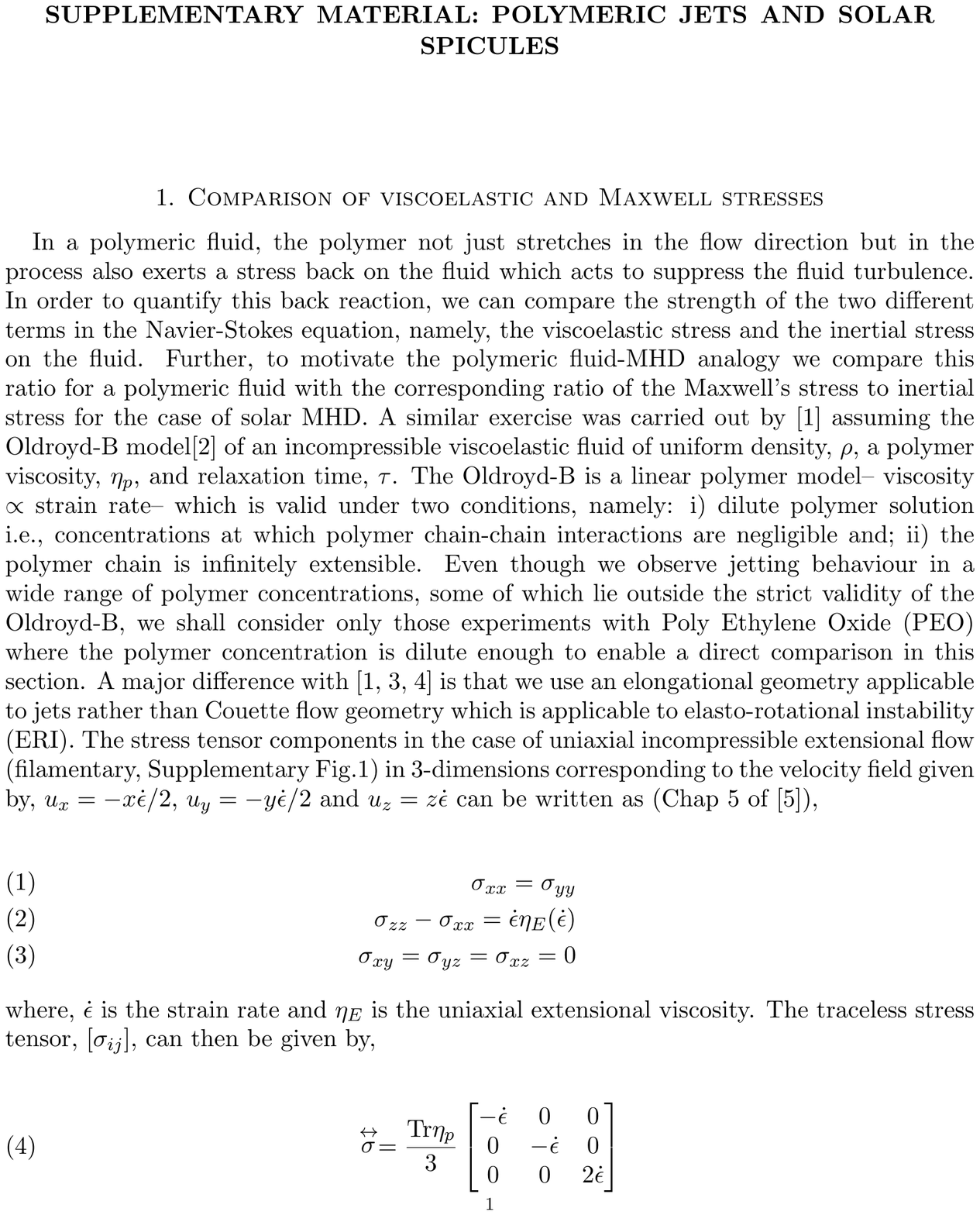}
\end{document}